\documentclass[aps,reprint]{revtex4-2}
\usepackage{etoolbox} 
\usepackage{graphicx}
\usepackage{subcaption}
\usepackage{multirow}%
\usepackage{amssymb}
\usepackage{xcolor}
\usepackage[table]{xcolor}
\usepackage{floatrow}
\usepackage{amsmath}
\usepackage[font=small,skip=3pt,justification=raggedright]{caption}
\usepackage{makecell}
\usepackage{textcomp}%
\usepackage{hyperref} 
\usepackage{cleveref}
\usepackage{url}
\usepackage{tikz}
\usepackage{siunitx}
\usepackage{enumitem}
\usepackage{booktabs}
\usepackage{array}
\usepackage[margin=1in]{geometry}
\DeclareSIUnit\year{yr}
\DeclareSIQualifier\thermal{th}
\usepackage{booktabs}
\usepackage{geometry}
\usepackage[version=4]{mhchem}
\usepackage{bm}
\let\mrm\mathrm

\makeatletter
\appto{\appendix}{%
  \@ifstar{\def\theequation@prefix{A.}}%
          {}
}
\makeatother

\newcolumntype{B}{>{\color{red}}c}



\begin{document}
\title{Production of High-Specific-Activity Radioisotopes Using High-Energy Fusion Neutrons}

\author{J. F. Parisi$^{1}$}
\email{jason@marathonfusion.com}
\author{A. Rutkowski$^{1}$}
\author{J. Harter$^{1,2}$}
\author{J. A. Schwartz$^{1}$}
\author{S. Chen$^{1,3}$}
\affiliation{$^1$Marathon Fusion, 150 Mississippi, San Francisco, 94107, CA, USA}
\affiliation{$^2$Department of Electronic \& Electrical Engineering, University of Bath, North Rd, Claverton Down, Bath, BA27AY, UK}
\affiliation{$^3$Stanford University, 450 Jane Stanford Way, Stanford, 94305, CA, USA}

\begin{abstract}
We show that transmutation driven by high-energy neutrons from deuterium–tritium (D–T) fusion reactions can produce many important medical radioisotopes—including $^{32}$P, $^{60}$Co, $^{64}$Cu, $^{89}$Sr, $^{90}$Y, $^{89}$Zr, $^{99}$Mo/$^{99\mathrm{m}}$Tc, $^{103}$Pd, $^{111}$In, $^{117}$In/$^{117\mathrm{m}1}$Sn, $^{123}$I, $^{125}$I, $^{131}$I, $^{133}$Xe, $^{153}$Sm, $^{166}$Ho, $^{177}$Lu, $^{188}$Re, and $^{192}$Ir—and emerging isotopes such as $^{47}$Sc, $^{67}$Cu, $^{103}$Ru/$^{103\mathrm{m}}$Rh, $^{103}$Pd/$^{103\mathrm{m}}$Rh, $^{119}$Sb, $^{124}$I, $^{155}$Tb, $^{161}$Tb, $^{195\mathrm{m}1}$Ir/$^{195\mathrm{m}}$Pt, and $^{225}$Ac with high specific activity and in large quantities. These reactions involve stable, abundant feedstocks and non-fission transmutation channels that change the proton number, enabling chemical separation of the product. Fusion-based transmutation could provide a flexible and proliferation-resistant platform for supply of high-purity isotopes. A D–T neutron source operating at a few megawatts of fusion power could meet or exceed global demand for most major radioisotopes. Further research is required to develop tailored approaches for feedstock processing and product extraction.
\end{abstract}

\maketitle

\section{Introduction}

In this Letter we report pathways to produce high-specific-activity (HSA) radioisotopes in large (grams to kilograms per megawatt year) quantities using neutrons from nuclear fusion reactions. Medical radioisotopes are indispensable in diagnostic imaging and targeted radionuclide therapy \cite{weiner1995radionuclides,yeong2014therapeutic}. Approximately 90\% of all nuclear medicine procedures rely on diagnostic isotopes \cite{EC2009MedicalRadioisotopes,WNA2025RadioIsotopes}, and among these, the ${}^{99}\mathrm{Mo}/{}^{99\mathrm{m}}\mathrm{Tc}$ generator pair accounts for most global demand. Yet production still depends on aging research reactors and complex international supply chains that have faced repeated interruptions \cite{OECDNEA2011_Supply,OECDNEA2010_Econ,NASEM_Mo99_2016,SNMMI2024_Mo99_Tc99m}. These vulnerabilities underscore the need for resilient, diversified, and HSA isotope production capabilities \cite{IAEA_TRS473_2011,habs2011production,hansen2022advancement}. We show that near-term fusion systems operating at the $\mathrm{MW_{th}}$ scale can meet this need. This capability motivates increased collaboration between the medical isotope, radiochemistry, and fusion energy communities to establish a near-term fusion volumetric neutron source for radioisotope production.

Deuterium–tritium (D–T) fusion, the leading candidate for near-term fusion plants \cite{Strachan1994short,Keilhacker1999,Wurzel2022}, produces 14.1 MeV neutrons,
\begin{equation}
  \mathrm{d} + \mathrm{t} \longrightarrow {}^{4}_{2}\mathrm{He} + {}^{1}_{0}\mathrm{n},
  \label{eq:DTreaction}
\end{equation}
whose energies exceed those of fission neutrons (around 1-2 MeV), opening new reaction channels such as $(\mathrm{n},2\mathrm{n})$, $(\mathrm{n},\mathrm{p})$, $(\mathrm{n},\alpha)$, and $(\mathrm{n},\mathrm{d})$. These reactions can efficiently produce HSA radioisotopes inaccessible by conventional methods.

Combined with chemical separation of product from feedstock, compact D–T neutron sources can enable scalable medical isotope production for a wide range of products. Promising candidates identified here include established and emerging isotopes $^{32}$P, $^{47}$Sc, $^{60}$Co, $^{64}$Cu, $^{67}$Cu, $^{89}$Sr, $^{90}$Y, $^{89}$Zr, $^{99}$Mo/$^{99\mathrm{m}}$Tc, $^{103}$Ru/$^{103\mathrm{m}}$Rh, $^{103}$Pd/$^{103\mathrm{m}}$Rh, $^{111}$In, $^{117}$In/$^{117\mathrm{m}1}$Sn, $^{119}$Sb, $^{123}$I,  $^{124}$I, $^{125}$I, $^{131}$I, $^{133}$Xe, $^{153}$Sm,  $^{155}$Tb, $^{161}$Tb,  $^{166}$Ho, $^{177}$Lu, $^{188}$Re, $^{192}$Ir, $^{195\mathrm{m}1}$Ir/$^{195\mathrm{m}}$Pt, and $^{225}$Ac. Earlier proposals for fusion-driven isotope production typically yielded low specific activity because they relied on $(\mathrm{n},\gamma)$ or (n,2n) reactions \cite{engholm1986radioisotope,Bourque1988FAME,Ridikas2006_HybridWaste,Leung2018_CompactNG,pietropaolo2021sorgentina,Honney2023FusionNeutrons,pereslavtsev2024potential}, with some HSA exceptions such as $\ce{^{176}Yb}(\mathrm{n},\gamma)\ce{^{177}Yb}$ for $\ce{^{177}Lu}$ \cite{ponsard2014production,lederer2024measurement}. In contrast, this work proposes compact, economically scalable systems using fusion-energy neutrons to drive transmutation that change the target’s proton number, allowing chemical separation to yield HSA radioisotopes. This work is complementary to recent studies of medical radioisotope production using D-T fusion neutrons \cite{li2023feasibility,evitts2025theoretical}, where our focus is wholly on transmutation pathways that enable production of HSA radioisotopes via elemental separation of the radioisotope from the feedstock.

\begin{figure}[bt!]
    \centering
    \begin{subfigure}[t]{0.99\textwidth}
    \centering
\includegraphics[width=0.9\textwidth]{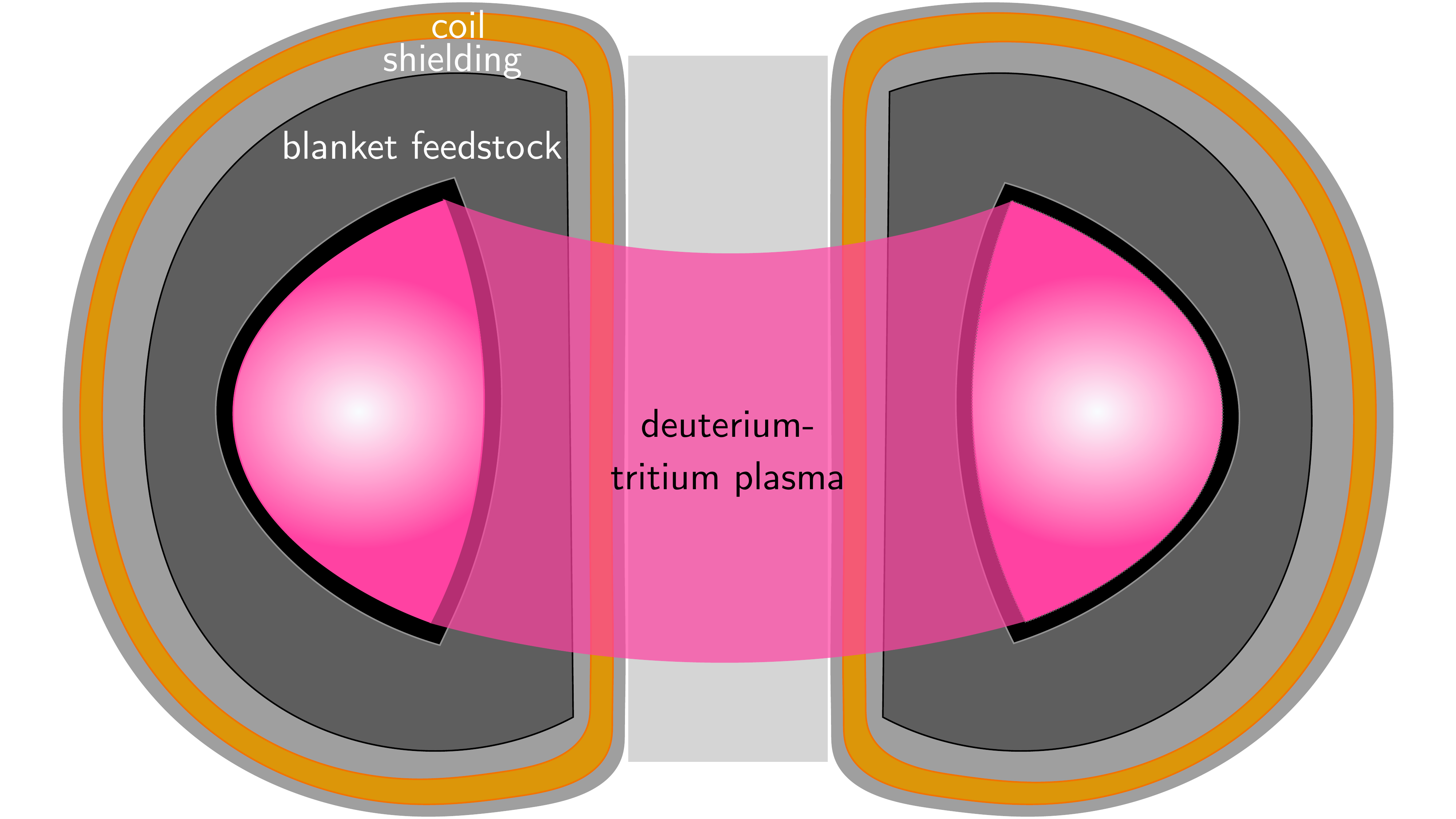}
    \end{subfigure}
    \caption{Schematic of an example toroidal D-T fusion neutron source with transmutation blanket.}
\label{fig:fusion_neutron_transmutation}
\end{figure}

This Letter is organized as follows: \Cref{sec:transmutation_basics} outlines the physics of fusion-neutron transmutation, \Cref{sec:transmutationpathways} presents key isotope pathways, and \Cref{sec:discussion} discusses broader implications.

\section{Fusion-Neutron Transmutation} \label{sec:transmutation_basics}

\begin{figure}[bt]
    \centering
    \begin{subfigure}[t]{0.99\textwidth}
    \centering
\includegraphics[width=1.0\textwidth]{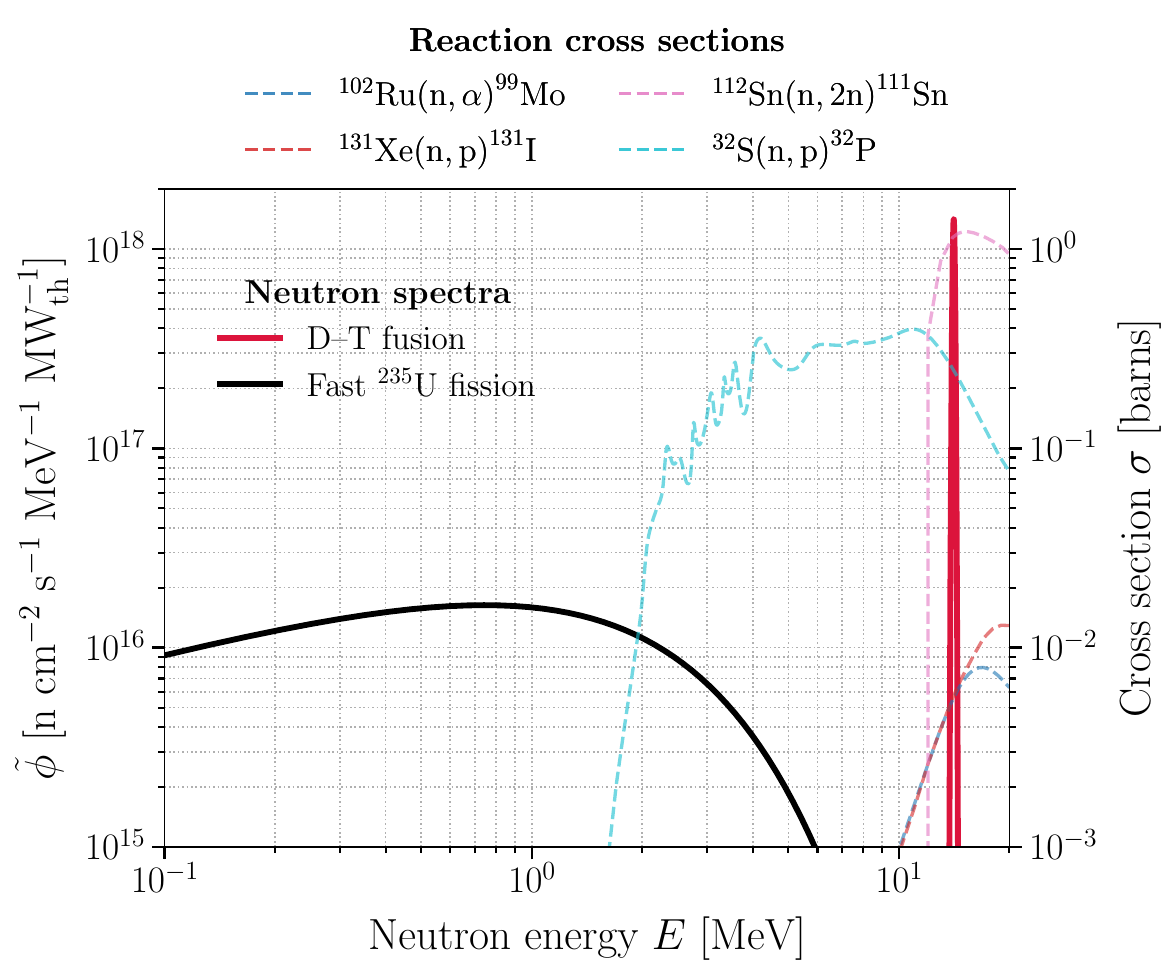}
    \caption{}
    \end{subfigure}
    \centering
    \begin{subfigure}[t]{0.99\textwidth}
    \centering
\includegraphics[width=1.0\textwidth]{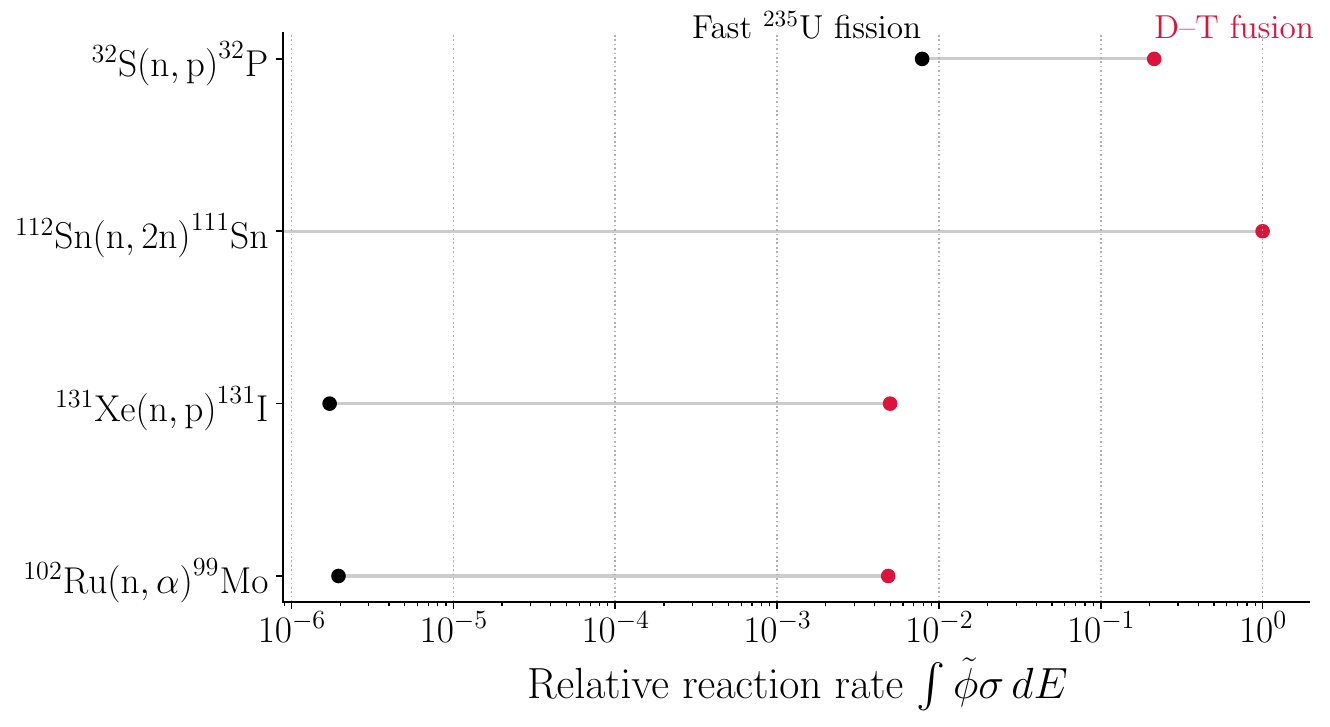}
    \caption{}
    \end{subfigure}
    \caption{(a) $\tilde{\phi}$ (\Cref{eq:phitilde}) for D-T fusion and fast $^{235}\mrm{U}$ neutron-birth spectra, reaction cross sections on secondary y-axis. (b) integrated relative reaction rate using neutron spectra in (a).}
\label{fig:production_rates}
\end{figure}

In a fusion system, D–T reactions produce 14.1~MeV neutrons that drive transmutation on a feedstock placed near the plasma, typically within a blanket region \cite{Rutkowski2025} (\Cref{fig:fusion_neutron_transmutation}). At these energies, threshold reactions such as $(\ce{n},2\ce{n})$, $(\mathrm{n},\mathrm{p})$, and $(\ce{n},\alpha)$ have cross sections large enough to yield substantial radioisotope production, and the absence of fission fragments enables simpler radioisotope extraction. The radioisotope production rate per unit volume is
\begin{equation}
\dot{n}_\mrm{p} = n_\mrm{t} \int \phi(E)\, \sigma(E)\, dE,
\label{eq:production_rate}
\end{equation}
where $n_\mrm{t}$ is the target number density, $\phi$ the neutron flux spectrum, $\sigma$ the reaction cross section, and $E$ the neutron energy. The resulting isotopes can in some cases be extracted using established chemical separation methods \cite{nichols1971status,maroni1975some,horwitz2005process,chattopadhyay2009simple,demange2016tritium,cristescu2020developments,holiski2024production}, though for many products new methods may need to be developed.

\begin{figure}[bt!]
    \centering
    \begin{subfigure}[tb]{1.\textwidth}
    \centering
    \includegraphics[width=1.0\textwidth]{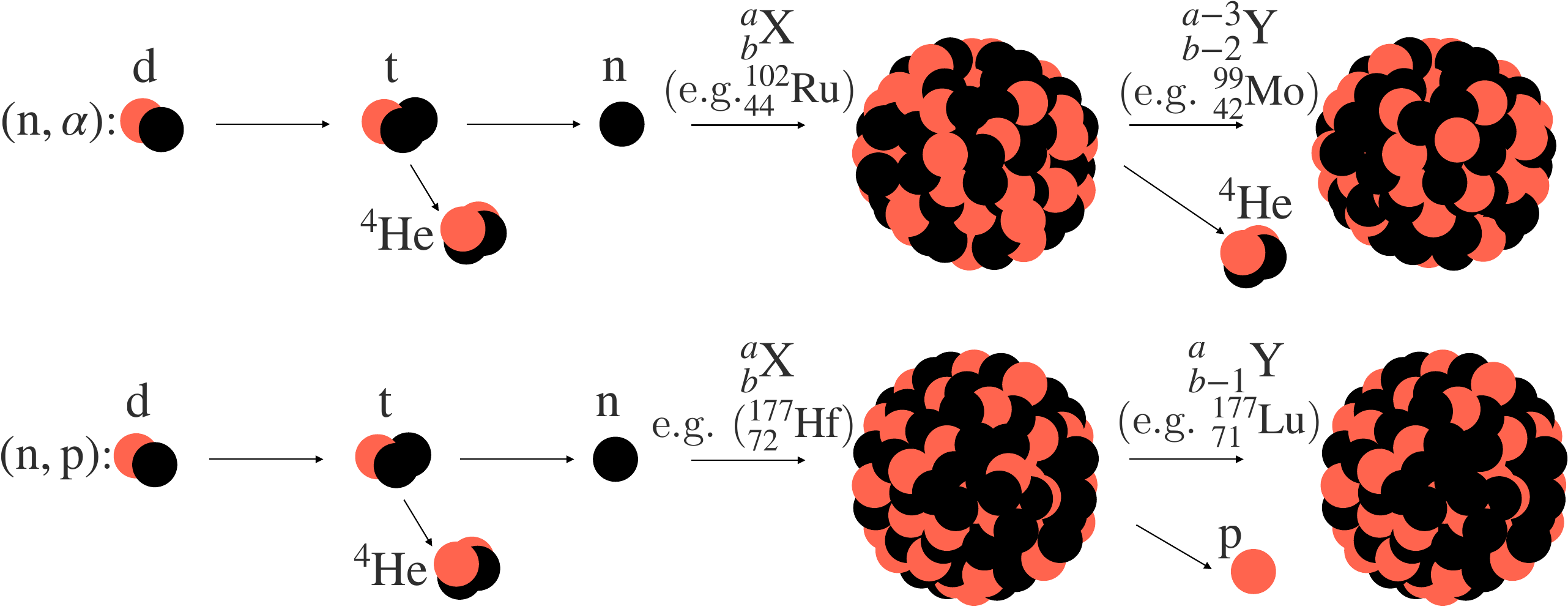}
    \end{subfigure}
    \caption{Graphical illustration of two transmutation pathways, $\ce{(n,\alpha)}$ and $\ce{(n,p)}$, driven by D-T neutrons.}
    \label{fig:transmutation_examples}
\end{figure}

It is instructive to compare transmutation rates achievable by fission and fusion-energy neutrons by considering the neutron flux per MW$_\mrm{th}$ of fission or fusion power,
\begin{equation}
    \tilde{\phi} \equiv \frac{\phi}{P} \nu_\mrm{excess},
    \label{eq:phitilde}
\end{equation}
where $P$ is the fission or fusion power in megawatts and $\nu_\mrm{excess}$ is the excess number of neutrons available for transmutation. We assume that fission releases $E_\mrm{fis} = 200$ MeV of energy with $\nu_\mrm{excess}=1.5$ (calculated using 2.5 neutrons per fission event, minus 1 neutron required to sustain the fission reactions) free neutrons per reaction \cite{terrell1962} and for fusion $\nu_\mrm{excess} = 1$. In \Cref{fig:production_rates}(a) we plot $\tilde{\phi}$ for an example $\ce{^235U}$ fast-fission Watt spectrum and D-T fusion neutron-birth spectrum - the fission spectrum has very few neutrons to drive $(\mathrm{n,p})$, $(\mathrm{n},\alpha)$, or $(\ce{n},2\ce{n})$ reactions. This is shown by the relative reaction rate $\int \tilde{\phi} \sigma dE$ in \Cref{fig:production_rates}(b), compared for fission and fusion. D-T fusion reactions drive many more $(\mathrm{n,p})$, $(\mathrm{n},\alpha)$, or $(\ce{n},2\ce{n})$ reactions per MW$_\mrm{th}$. Illustrations of $\ce{(n,\alpha)}$ and $\ce{(n,p)}$ transmutations are shown in \Cref{fig:transmutation_examples}. A shortcoming of using the neutron birth-energy spectra is that it omits the effects of neutrons damping in materials -- in the next section we perform neutron transport simulations to capture these effects.

\begin{figure*}[tb!]
    \centering
    \includegraphics[width=1.0\textwidth]{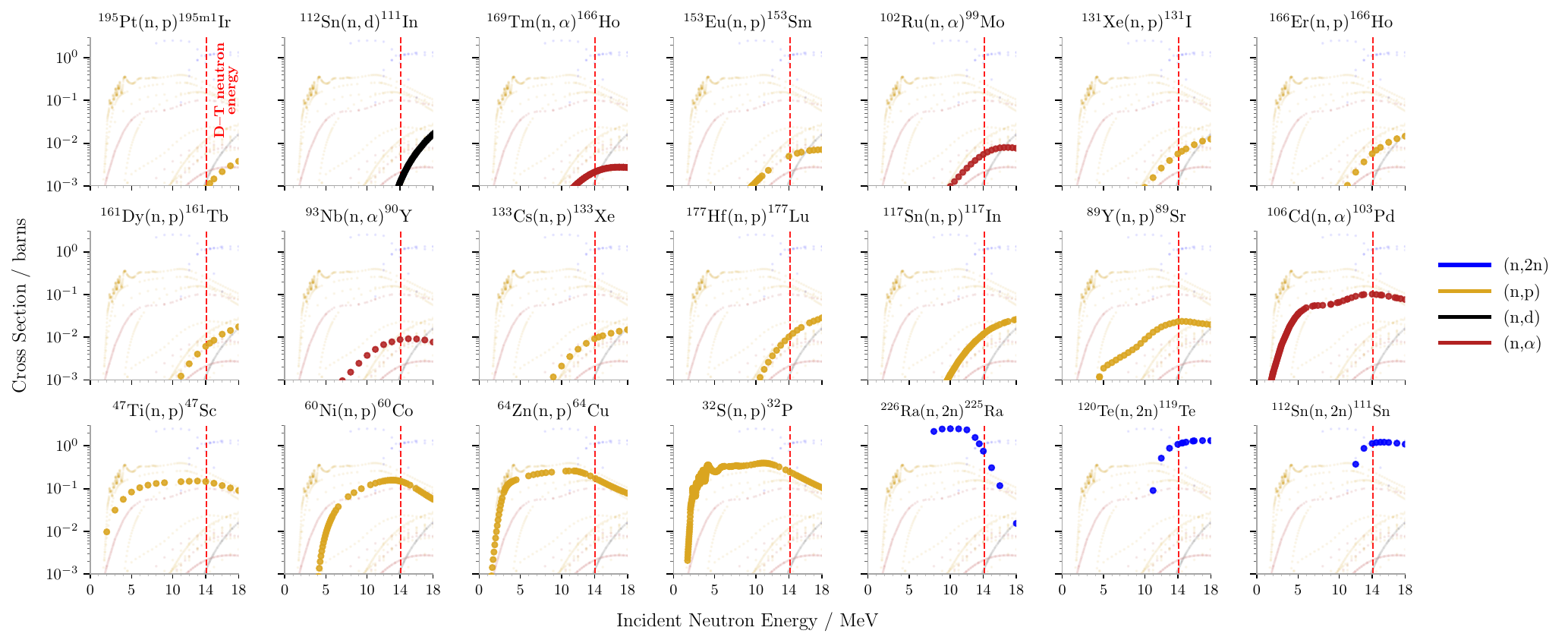}
    \caption{Cross sections corresponding to transmutation pathways for radioisotopes. Data from \cite{Brown20181}.}
    \label{fig:transmutation_pathways}
\end{figure*}

For the fusion-neutron-driven reactions considered here, $\sigma$ spans from millibarns to barns, shown in \Cref{fig:transmutation_pathways}. Typical fusion blankets have a neutron spectrum peaked at 14.1 MeV, with flux decreasing at lower energies as neutrons slow through scattering \cite{gorley2020eurofusion}. Because medical isotopes usually require HSA, separability from the feedstock is essential. Thus, reactions that change the target’s proton number---such as $(\mathrm{n,p})$ or $(\mathrm{n},\alpha)$---are preferred, whereas $(\ce{n},2\ce{n})$ and $(\ce{n},\gamma)$ routes generally (but not always, if appropriate decay pathways exist) produce isotopic mixtures where the vast majority of the mixture is the feedstock isotope mix. Although $(\mathrm{n,p})$ and $(\mathrm{n},\alpha)$ cross sections are generally much smaller than (n,2n), MW-scale fusion systems employing $(\mathrm{n,p})$ and $(\mathrm{n},\alpha)$ pathways can still easily meet or exceed global demand for most radioisotopes.

Fusion-driven transmutation offers several advantages over fission and accelerator-based methods. D–T fusion produces roughly an order of magnitude more neutrons than fission for each watt of power, and has minimal long-lived waste. Existing approaches based on fission also require the use of $\mathrm{^{235}U}$ - often enriched - resulting in additional regulatory challenges for these systems \cite{NNSA_2023_Mo99_domestic_supply}. The absence of fission products simplifies chemical separation, while fusion provides higher neutron rate, neutron flux, and energy efficiency compared with accelerators \cite{schmor2011review,qaim2012present,starovoitova2014production,wang2022production}. Together, these features make fusion-neutron transmutation an attractive, proliferation-resistant platform for medical isotope production.

To quantify isotope yields and specific activity, we perform neutron transport simulations with depletion using OpenMC \cite{romano2015openmc}. A 14.1 MeV neutron source bombards a 10$\times$10 cm slab of thickness 20 cm with a flux of $10^{14}$ neutrons / (cm$^2$ second). This corresponds to a neutron wall loading of $\sim 2.3\mathrm{MW/m^2}$, typical with design targets for fusion power systems \cite{wong2000tokamak,bolt2004materials,lyon2008systems,meier2014fusion,Sorbom2015}.

\section{Transmutation Pathways} \label{sec:transmutationpathways}

\begin{figure*}[tb!]
    \centering
    \begin{subfigure}[t]{0.39\textwidth}
    \centering \includegraphics[width=1.0\textwidth]{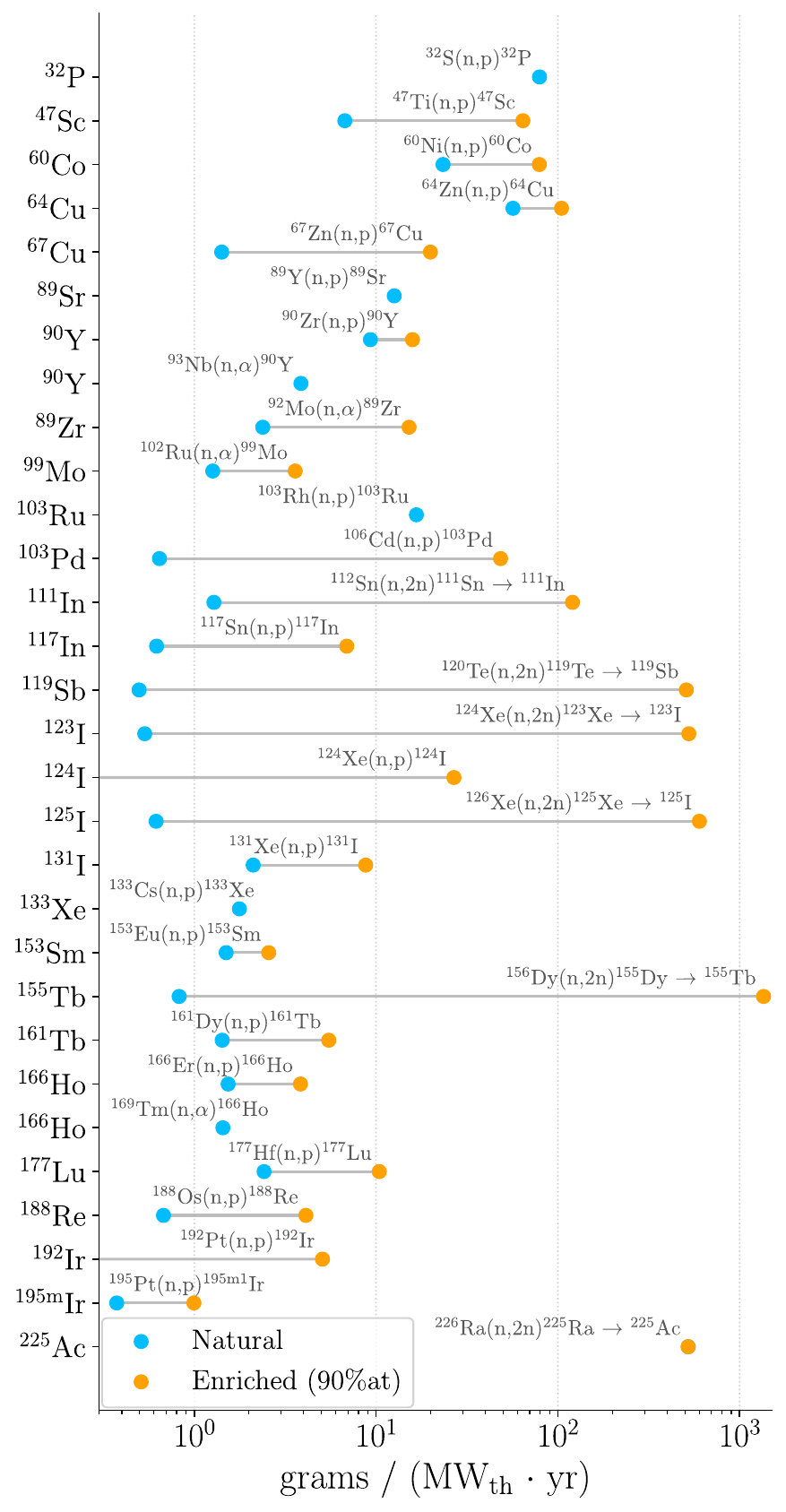}
    \caption{}
    \end{subfigure}
    \centering
    \begin{subfigure}[t]{0.55\textwidth}
    \centering \includegraphics[width=0.99\textwidth]{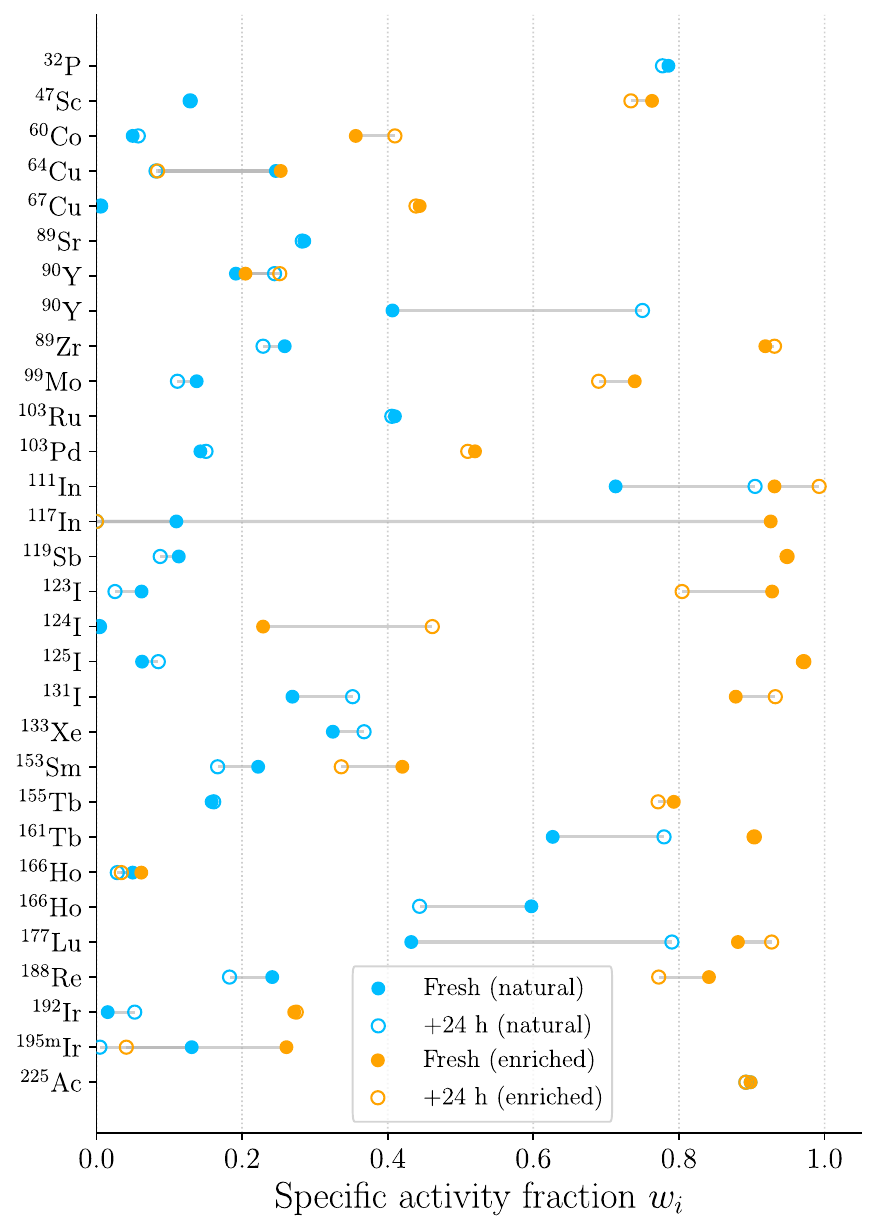}
    \caption{}
    \end{subfigure}
    \caption{(a) Radioisotope production in grams per MWth year, (b) mass fraction $w_i$ (see \Cref{eq:wi}) in a D-T neutron flux of $10^{14}$n/s/cm$^2$($\sim2.3~\mathrm{MW/m^2}$ neutron loading). In (b), `+24 h' corresponds to extracted radioisotope material 24 hours after initial extraction.}
    \label{fig:product_output}
\end{figure*}

In this section, we present transmutation pathways for HSA radioisotope production in fusion blankets using neutronics simulations. For the transmutation pathways presented here, we only considered feedstocks that are stable and abundant, ensuring that our proposed production pathways are practically realizable. The one exception to this rule is the use of $^{226}$Ra as feedstock for production of $^{225}$Ac, given that this feedstock is known to be available in gram-scale quantities. The main results are summarized in \Cref{tab:medical_isotopes_with_number_density}.

Production rates in \qty{}{\gram\per\MW\thermal\year} for all pathways are shown in \Cref{fig:product_output}(a) for natural and enriched feedstock - given that global demand for most medical radioisotopes is at most several grams per year, a several MW${}_\mrm{th}$ fusion source could supply the entire demand for almost any single radioisotope considered, using unenriched feedstock. With enriched feedstock, a 10 MW${}_\mrm{th}$ fusion source could supply almost all of the listed radioisotopes.

A crucial quantity is the specific activity of the target radioisotope, normalized by the mass of all isotopes in the mixture,
\begin{equation}
    \overline{A}_s \equiv \frac{\lambda_i}{m_\mrm{tot}} = \lambda_i w_i \frac{ N_\mrm{A}}{M_i},
    \label{eq:A_overline}
\end{equation}
where $\lambda_i$ is the activity of the target isotope, $w_i$ is the mass fraction of target radioisotope,
\begin{equation}
    w_i \equiv \frac{m_i}{m_\mrm{tot}},
    \label{eq:wi}
\end{equation}
$m_i$ is the target isotope mass, $m_\mrm{tot}$ is the mass of all isotopes in the mixture, $N_\mrm{A}$ is Avogadro's constant, and $M_i$ is the molar mass of the target radioisotope. In order to calculate $w_i$ from neutronics simulations, we extract an elementally pure mixture of the target radioisotope every 30 minutes for one hour (except for $\ce{^111In}$, extracted every 5 minutes for 10 minutes due to short $\ce{^111Sn}$ half-life). We then perform continuous elemental separation on the mixture for 24 hours (the assumed time between extraction and delivery), removing all non-target elements resulting from radioactive decay. We measure $w_i$ both at fresh extraction and 24 hours later - shown in \Cref{fig:product_output}(b). 

\begin{figure}[bt!]
    \centering
    \begin{subfigure}[tb]{1.\textwidth}
    \centering
    \includegraphics[width=1.0\textwidth]{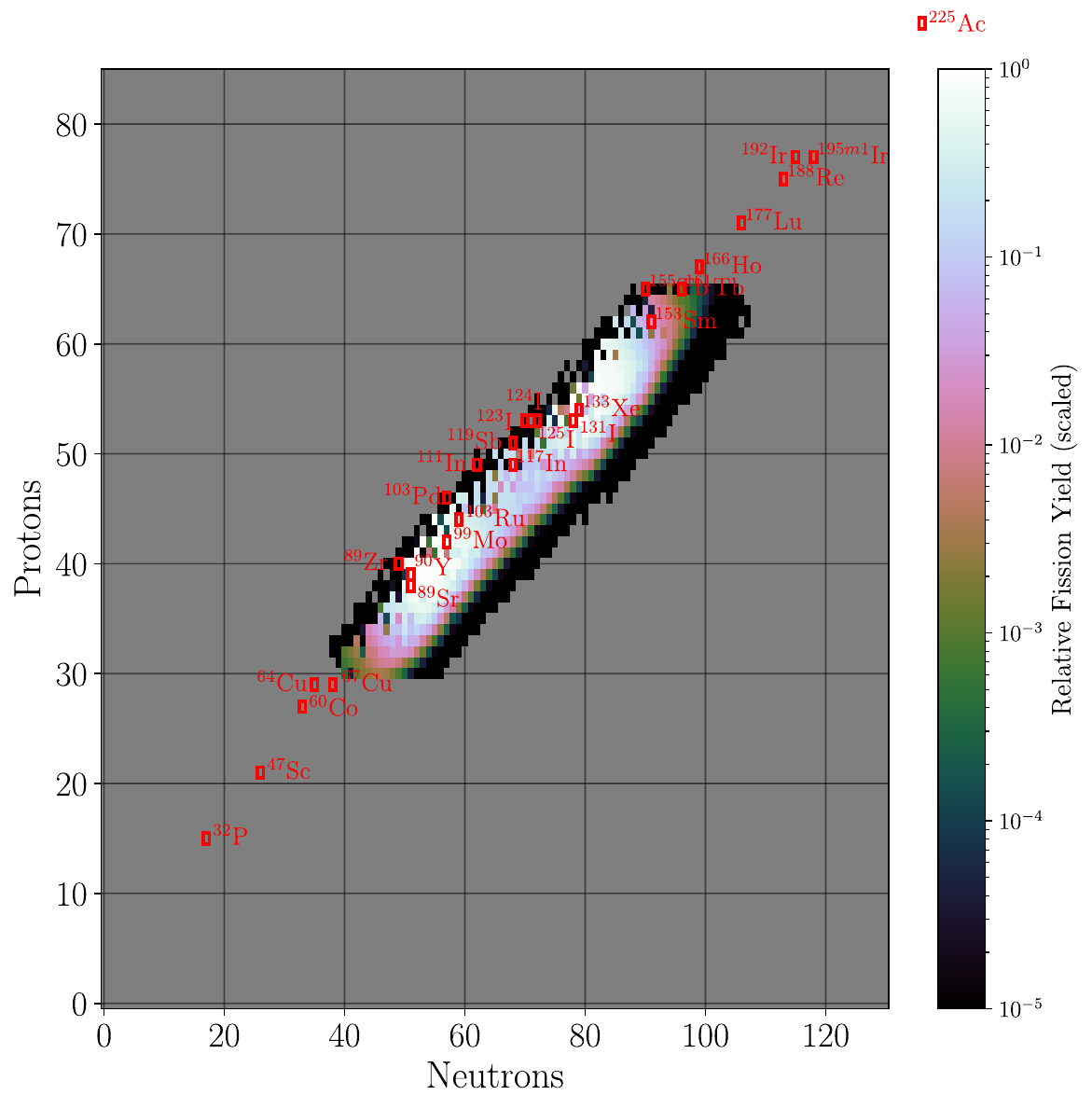}
    \caption{}
    \end{subfigure}
    \centering
    \begin{subfigure}[tb]{1.\textwidth}
    \centering
    \includegraphics[width=1.0\textwidth]{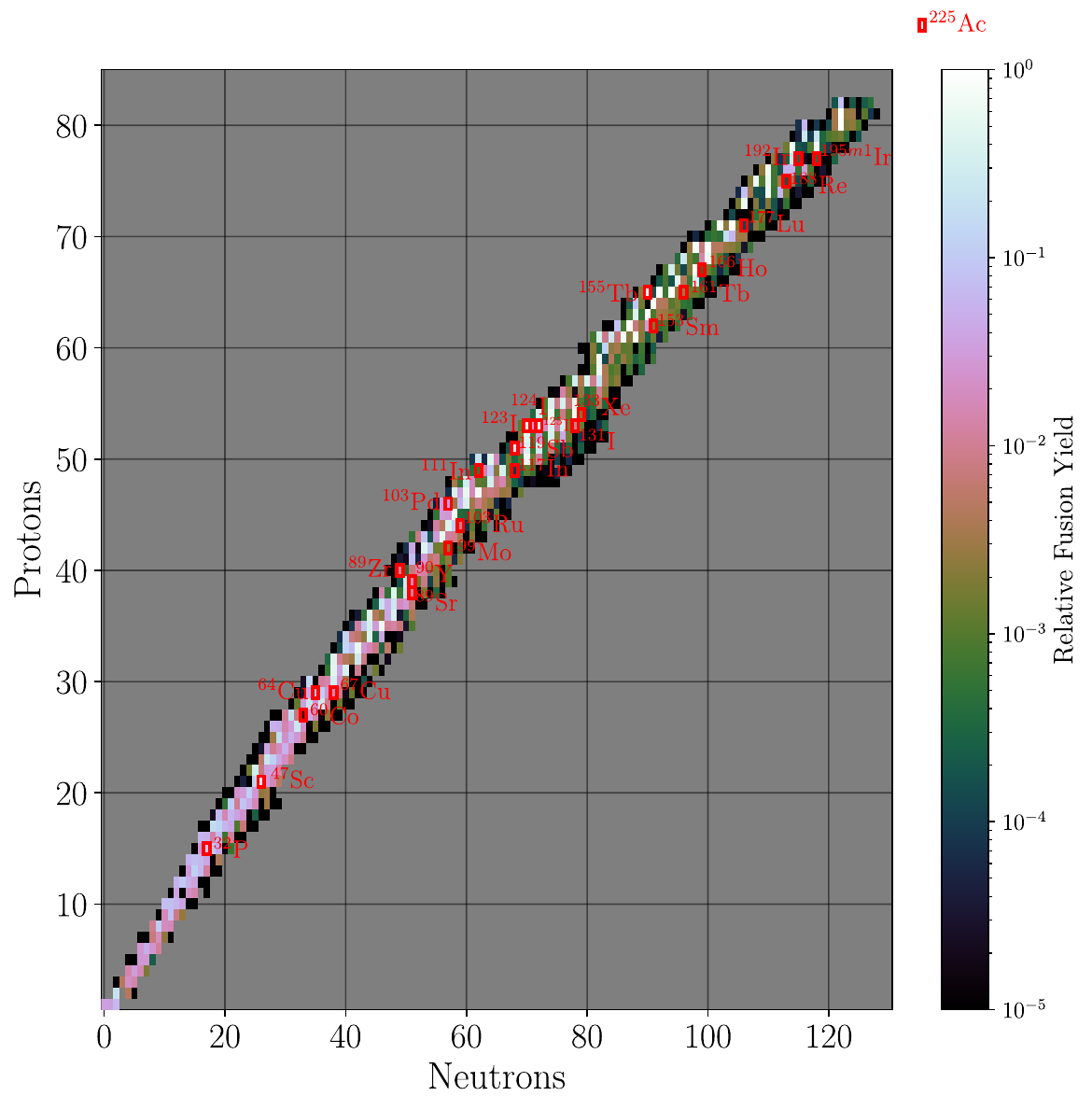}
    \caption{}
    \end{subfigure}
    \caption{(a) $\ce{^235U}$ fission product yields with radioisotopes considered in this paper highlighted with red boxes. (b) Relative fusion yields for all target isotopes using fusion neutrons; only neutron-driven transmutations that change proton number of feedstock to product are considered.}
    \label{fig:transmutation_space}
\end{figure}

Therapeutic $\beta^-$ emitters such as ${}^{32}$P, ${}^{89}$Sr, ${}^{90}$Y, ${}^{153}$Sm, ${}^{166}$Ho, ${}^{177}$Lu, and ${}^{188}$Re are produced through $(\ce{n,p})$ or $(\ce{n,\alpha})$ reactions, with typical outputs from 1 to \qty{80}{\gram\per\MW\thermal\year} for natural feedstocks. 
${}^{32}$P is long established for hematologic and labeling applications \cite{IAEA_INIS_P32_2000,INIS_P32_Assessment_2013,Azarov2021_P32_NG}, while ${}^{89}$Sr and ${}^{153}$Sm are used for metastatic bone pain \cite{Pantel2023,Dash2025,Voorde2021_Frontiers_Sm153HSA,Vermeulen2022_Sm153_MassSep}. 
${}^{90}$Y and ${}^{177}$Lu underpin radioligand and radioembolization therapies \cite{Weber2022,Tennvall2007,Basu2020,Vogel2021_Lu177_Review,Morris2025_CurrMedChem_DeuteronLu177}, and emerging emitters such as ${}^{166}$Ho and ${}^{161}$Tb offer improved dosimetry and imaging compatibility \cite{Vosoughi2017_IJNMM_Dy_Ho,Zimmerman2024_DDEP_166Ho,Prince2018_JNM_Ho166,Klaassen2019_EJNMMIPharmChem_Ho166,Ntihabose2025_EJNMMIPharmChem_Tb161Review,Santos2025_Tb161_Mito}. 
${}^{188}$Re from $\ce{^{188}Os(n,p)^{188}Re}$ is a direct analogue to existing generator systems \cite{Kondev2018}.

Diagnostic and theranostic isotopes including ${}^{64}$Cu, ${}^{111}$In, ${}^{131}$I, ${}^{195\mathrm{m}}$Pt, and ${}^{103\mathrm{m}}$Rh have various functions. ${}^{64}$Cu provides dual PET and therapeutic capability \cite{Zhou2019_Cu64_Review,Svedjehed2020_EJNMMIPharmChem,IAEA_2019_Cu64_Brochure,Krasnovskaya2023_Cu64_67_Review}, while ${}^{111}$In and ${}^{131}$I remain clinical standards for SPECT imaging and thyroid therapy \cite{Mohammed2020_AIP_IndiumRoutes,Indium111_TopicPage_Elsevier,Avram2022_JNM_Guideline,Petranovic2022_EJNMMI_ETA,Mishra2021_ARI_131IYield,Chattopadhyay2009_ARI_131I_Separation}. 
${}^{195\mathrm{m}}$Pt, produced via the short-lived ${}^{195\mathrm{m1}}$Ir isomer, has growing interest as a platinum-based theranostic \cite{Hilgers2008Pt195mAlpha,Knapp2005_HFIR_Isotopes,Madumarov2024_Pt195m_DoubleCap,deRoest2024_Pt195m_PD,Hoogenkamp2025_Pt195m_Clinical}, while ${}^{103\mathrm{m}}$Rh is an Auger emitter obtainable from ${}^{103}$Ru or ${}^{103}$Pd generators \cite{Saidi2011_Pd103_Rhpn,Uncu2023_Pd103_Routes,Hindie2024_Pd103_Dosimetry}.

Among generator systems, ${}^{99}$Mo/${}^{99\mathrm{m}}$Tc remains the global clinical standard \cite{harper1965technetium,eckelman2009unparalleled,papagiannopoulou2017technetium}, with the fusion-accessible $\ce{^{102}Ru(n,\alpha)^{99}Mo}$ reaction \cite{gascoine2021towards} providing a proliferation-resistant route at \qty{1.3}{\gram\per\MW\thermal\year} for natural ruthenium and \qty{3.6}{\gram\per\MW\thermal\year} for 90\%at enriched $\ce{^102Ru}$; after 24 hours, the enriched path provides a mixture containing $\sim69\%$at ${}^{99}$Mo. We contrast the $\ce{^{102}Ru(n,\alpha)^{99}Mo}$ approach here with previous work favoring $\ce{^{98}Mo(n,\gamma)^{99}Mo}$ and $\ce{^{100}Mo(n,\mrm{2}n)^{99}Mo}$ pathways \cite{pietropaolo2021sorgentina,li2023feasibility,evitts2025theoretical}, which resulted in much higher absolute production (in grams) of ${}^{99}$Mo, but much lower specific activity. \cite{evitts2025theoretical} found $\ce{^{100}Mo(n,\mrm{2}n)^{99}Mo}$ gave a specific activity of $\approx 14$Ci/g at much higher ${}^{99}$Mo production quantities, whereas we find $\ce{^{102}Ru(n,\alpha)^{99}Mo}$ gives specific activity of $3.3\cdot10^5$Ci/g following chemical separation of ${}^{99}$Mo from $\ce{^102Ru}$ and subsequent cooling. Therefore, the $\ce{^102Ru}$ pathway is preferable for HSA ${}^{99}$Mo and the $\ce{^98Mo}$, $\ce{^100Mo}$ pathways for low-specific-activity ${}^{99}$Mo.

Alpha emitters have significant therapeutic potential \cite{poty2018alpha,poty2018alphapt2} - in this work we only consider producing $\ce{^225Ac}$ from $\ce{^226Ra}$ feedstock - more alpha emitters produced by fusion neutrons are considered in \cite{evitts2025theoretical}. Our simulations indicate that under a neutron flux of \qty{e14}{\per\centi\meter\squared\second}, roughly 1.3g of $\ce{^225Ac}$ is produced per kg of $\ce{^226Ra}$ feedstock when the $\ce{^226Ra}$ layer is 20cm deep. For layers of ~2cm thickness, around 3g of $\ce{^225Ac}$ is produced per kg of $\ce{^226Ra}$ feedstock. Given that $\ce{^226Ra}$ is scarce and expensive \cite{iyengar1990natural}, consider a smaller system containing $\sim$11g of $\ce{^226Ra}$ in an oblong with surface area 1cm$^2$ and 2cm deep. A \qty{e14}{\per\centi\meter\squared\second} neutron flux would produce $\sim$0.033g of $\ce{^225Ac}$ / year, roughly 650 times larger than present (2025) global supply and would require approximately 280 Watts of fusion neutrons passing through the surface.

Given the wide range of feedstock compositions and product chemistries considered here, each transmutation pathway will likely require a dedicated extraction system optimized for the product’s volatility, chemical form, and half-life. In general, the extraction strategy depends on whether the product remains in the solid, liquid, or gaseous phase under blanket conditions. 

\sisetup{
exponent-product=\cdot,
}

\begin{table*}[htp]
\centering
\footnotesize
\setlength{\tabcolsep}{2.2pt}
\renewcommand{\arraystretch}{0.9}
\caption{Annual medical radioisotope production driven by D-T fusion neutrons with (1) natural abundance and (2) a 90\%at enriched blanket at a neutron flux of \qty{e14}{\per\centi\meter\squared\second} and with a feedstock thickness of \qty{20}{cm}. The parameter $w_i$ denotes the fractional inventory remaining at start (Fresh) and after 24 hours of cooldown. For generator pairs such as $^{99}$Mo/$^{99\mathrm{m}}$Tc, all quantities correspond to the generator ($^{99}$Mo), not the daughter isotope. $^\dagger$For $\ce{^225Ac}$ we use 100\% $\ce{^226Ra}$ feedstock.}
\begin{tabular}{@{}r@{}c c c c cccc cc ccc@{}}
\toprule
\multicolumn{2}{c}{\thead{Radio- \\ isotope}} &
\thead{Half-life \\ (days)} &
\thead{Transmutation \\ Pathway(s)} &
\thead{Feedstock \\ Isotopic \\ Abundance \\ (\%)} &
\multicolumn{2}{c}{$w_i$ (Fresh)} &
\multicolumn{2}{c}{$w_i$ (24\,h)} &
\multicolumn{2}{c}{\thead{Fusion prod. \\ (g\,/(MW$_\mathrm{th}\,\mrm{yr}$))}} &
\multicolumn{2}{c}{\thead{Feedstock \\ Num. Density \\ (\qty{e20}{\per\cubic\centi\meter})}} &
\thead{Maturity} \\ 
\cmidrule(lr){6-7} \cmidrule(lr){8-9} \cmidrule(lr){10-11} \cmidrule(lr){12-13}
&&&&& \thead{nat.} & \thead{enr.} & \thead{nat.} & \thead{enr.} & \thead{nat.} & \thead{enr.} & \thead{nat.} & \thead{enr.} & \\ 
\arrayrulecolor[gray]{0.85}\hline

 & $^{32}$P & 14.3 & $\ce{^{32}S(n,p)^{32}P}$ & 95 & 0.79 & -- & 0.78 & -- & 79 & -- & 370 & -- & \makecell[c]{FDA-appr. \\ (hematologic)} \\ \hline
 & $^{47}$Sc & 3.35 & $\ce{^{47}Ti(n,p)^{47}Sc}$ & 7.4 & 0.13 & 0.76 & 0.13 & 0.73 & 6.7 & 64 & 42 & 510 & \makecell[c]{Emerging \\ (theranostic, \\ preclinical)} \\ \hline
 & $^{60}$Co & 1925 & $\ce{^{60}Ni(n,p)^{60}Co}$ & 26 & 0.0498 & 0.36 & 0.057 & 0.41 & 23 & 79 & 235 & 810 & \makecell[c]{FDA-appr. \\ (radiotherapy)} \\ \hline
 & $^{64}$Cu & 0.53 & $\ce{^{64}Zn(n,p)^{64}Cu}$ & 49 & 0.25 & 0.25 & 0.082 & 0.084 & 57 & 100 & 327 & 600 & \makecell[c]{Trials \\ (PET agent)} \\ \hline
 & $^{67}$Cu & 2.58 & $\ce{^{67}Zn(n,p)^{67}Cu}$ & 4 & 0.006 & 0.006 & 0.44 & 0.44 & 1.4 & 20 & 26 & 579 & \makecell[c]{Trials \\ (theranostic \\ radiotherapy)} \\ \hline
 & $^{89}$Sr & 50.6 & $\ce{^{89}Y(n,p)^{89}Sr}$ & 100 & 0.29 & -- & 0.28 & -- & 13 & -- & 303 & -- & \makecell[c]{FDA-appr. \\ (metastasis)} \\ \hline
 & $^{89}$Zr & 3.27 & $\ce{^{92}Mo(n,\alpha)^{89}Zr}$ & 15 & 0.26 & 0.92 & 0.23 & 0.93 & 2.4 & 15 & 101 & 606 & \makecell[c]{Trials \\ (immuno-PET)} \\ \hline
 & $^{90}$Y & 2.67 & \makecell[c]{$\ce{^{90}Zr(n,p)^{90}Y}$ \\ $\ce{^{93}Nb(n,\alpha)^{90}Y}$} & \makecell[c]{52 \\ 100} & \makecell[c]{0.19 \\ 0.41} & \makecell[c]{0.20 \\ --} & \makecell[c]{0.24 \\ 0.75} & \makecell[c]{0.25 \\ --} & \makecell[c]{3.9 \\ 9.3} & \makecell[c]{8.8 \\ 16} & \makecell[c]{224 \\ 555} & \makecell[c]{390 \\ 500} & \makecell[c]{FDA-appr. \\ (microspheres)} \\ \hline
 & $^{99}$Mo/$^{99\mathrm{m}}$Tc & 2.75 & $\ce{^{102}Ru(n,\alpha)^{99}Mo}$ & 32 & 0.14 & 0.74 & 0.11 & 0.69 & 1.3 & 3.6 & 229 & 640 & \makecell[c]{FDA-appr. \\ (clinical standard)} \\ \hline
 & $^{103}$Ru/$^{103\mathrm{m}}$Rh & 39.2 & $\ce{^{103}Rh(n,p)^{103}Ru}$ & 100 & 0.41 & -- & 0.41 & -- & 17 & -- & 730 & -- & \makecell[c]{pre-clinical} \\ \hline
 & $^{103}$Pd/$^{103\mathrm{m}}$Rh & 17.0 & $\ce{^{106}Cd(n,p)^{103}Pd}$ & 1.3 & 0.14 & 0.52 & 0.15 & 0.51 & 0.64 & 48 & 6.2 & 430 & \\ \hline
 & $^{111}$In & 2.80 & \makecell[c]{$\ce{^{112}Sn(\mrm{n,2n})^{111}Sn}$ \\ $\ce{->[{\beta^+}][{35\ \text{min}}] ^{111}In}$} & 1.0 & 0.71 & 0.93 & 0.90 & 0.99 & 1.3 & 120 & 3.8 & 340 & \makecell[c]{FDA-appr. \\ (diagnostic)} \\ \hline
 & $^{117}$In/$^{117\mathrm{m}1}$Sn & 0.03 & $\ce{^{117}Sn(n,p)^{117}In}$ & 7.7 & 0.11 & 0.93 & 0.0 & 0.0 & 0.62 & 6.9 & 29 & 340 & \makecell[c]{FDA-appr. \\ (palliation)} \\ \hline
 & $^{119}$Sb & 1.59 & \makecell[c]{$\ce{^{120}Te(\mrm{n,2n})^{119}Te}$ \\ $\ce{->[\mrm{EC}(98\%),\beta^+(2\%)][{16\ \text{hr}}] ^{119}Sb}$} & 0.09 & 0.11 & 0.95 & 0.09 & 0.95 & 0.50 & 509 & 0.30 & 300 & \makecell[c]{pre-clinical} \\ \hline
  & $^{123}$I & 0.55 & \makecell[c]{$\ce{^{124}Xe(\mrm{n,2n})^{123}Xe}$ \\ $\ce{->[{\beta^+}][{2.1\ \textrm{hr}}] ^{123}I}$} & 0.10 & 0.062 & 0.93 & 0.025 & 0.80 & 0.53 & 525 & 0.14 & 128 & \makecell[c]{FDA-appr. \\ (SPECT thyroid \\ neuro)} \\ \hline 
 & $^{124}$I & 4.18 & $\ce{^{124}Xe(n,p)^{124}I}$ & 0.10 & 0.0035 & 0.23 & 0.0042  & 0.46 & 0.027 & 27 & 0.14 & 128 & \makecell[c]{Trials \\ (PET imaging)} \\ \hline
 & $^{125}$I & 59.4 & \makecell[c]{$\ce{^{126}Xe(\mrm{n,2n})^{125}Xe}$ \\ $\ce{->[{\beta^+}][{16.9\ \textrm{hr}}] ^{125}I}$} & 0.09 & 0.063 & 0.97 & 0.085  & 0.97 & 0.62 & 601 & 0.12 & 126 & \makecell[c]{FDA-appr. \\ (brachytherapy)} \\ \hline 
 & $^{131}$I & 8.02 & $\ce{^{131}Xe(n,p)^{131}I}$ & 21 & 0.27 & 0.88 & 0.35 & 0.93 & 2.1 & 8.8 & 29 & 120 & \makecell[c]{FDA-appr. \\ (thyroid)} \\ \hline
 & $^{133}$Xe & 5.25 & $\ce{^{133}Cs(n,p)^{133}Xe}$ & 100 & 0.32 & -- & 0.37 & -- & 1.8 & -- & 88 & -- & \makecell[c]{FDA-appr. \\ (pulmonary)} \\ \hline
 & $^{153}$Sm & 1.93 & $\ce{^{153}Eu(n,p)^{153}Sm}$ & 52 & 0.22 & 0.42 & 0.17 & 0.34 & 1.5 & 2.6 & 108 & 190 & \makecell[c]{FDA-appr. \\ (bone-pain)} \\ \hline
 & $^{155}$Tb &5.23 & \makecell[c]{$\ce{^{156}Dy(\mrm{n,2n})^{155}Dy}$ \\ $\ce{->[{\beta^+}][{9.9\ \textrm{hrs}}]^{155}Tb}$} & 0.056 & 0.16  & 0.79 & 0.16 & 0.77 & 0.83 & 1354 & 0.18 & 297 & \makecell[c]{Emerging \\ (SPECT theranostic)} \\ \hline
 & $^{161}$Tb & 6.95 & $\ce{^{161}Dy(n,p)^{161}Tb}$ & 19 & 0.63 & 0.90 & 0.78 & 0.90 & 1.4 & 5.5 & 60 & 280 & \makecell[c]{Trials \\ (alt.\ to ${}^{177}\mathrm{Lu}$)} \\ \hline
 & $^{166}$Ho & 1.12 & \makecell[c]{$\ce{^{166}Er(n,p)^{166}Ho}$ \\ $\ce{^{169}Tm(n,\alpha)^{166}Ho}$} & \makecell[c]{33 \\ 100} & \makecell[c]{0.050 \\ 0.60} & \makecell[c]{0.062 \\ --} & \makecell[c]{0.029 \\ 0.44} & \makecell[c]{0.034 \\ --} & \makecell[c]{1.5 \\ 1.4} & \makecell[c]{3.8 \\ --} & \makecell[c]{110 \\ 332} & \makecell[c]{300 \\ --} & \makecell[c]{CE-marked in Europe \\ (clinical trials, \\ not FDA-appr.)} \\ \hline
 & $^{177}$Lu & 6.64 & $\ce{^{177}Hf(n,p)^{177}Lu}$ & 19 & 0.43 & 0.88 & 0.79 & 0.93 & 2.4 & 10 & 84 & 400 & \makecell[c]{FDA-appr. \\ (radioligand)} \\ \hline
 & $^{188}$Re & 0.708 & $\ce{^{188}Os(n,p)^{188}Re}$ & 13 & 0.24 & 0.84 & 0.18 & 0.77 & 0.68 & 4.1 & 96 & 660 & \makecell[c]{FDA-appr. \\ (generator)} \\ \hline
 & $^{192}$Ir & 73.8 & $\ce{^{192}Pt(n,p)^{192}Ir}$ & 0.78 & 0.015 & 0.27 & 0.053  & 0.27 & 0.044 & 5.1 & 5.3 & 606 & \makecell[c]{FDA-appr. \\ (HDR brachytherapy)} \\ \hline
 & $^{195\mathrm{m}1}$Ir/$^{195\mathrm{m}}$Pt & 0.156 & $\ce{^{195}Pt(n,p)^{195\mrm{m}1}Ir}$ & 34 & 0.13 & 0.26 & 0.0048 & 0.041 & 0.37 & 1.0 & 223 & 590 & \makecell[c]{Pilot \\ (pre-clinical)} \\ 
 & $^{225}$Ac & 9.92 & \makecell[c]{$\ce{^{226}Ra(\mrm{n,2n})^{225}Ra}$ \\ $\ce{->[\beta^-][{15\ \text{days}}] ^{225}Ac}$} & Trace$^\dagger$ & 0.90 & -- & 0.98 & -- & 521 & -- & 132 & -- & \makecell[c]{Trials \\ (alpha therapy)} \\ \hline
\arrayrulecolor{black}\hline
\end{tabular}
\label{tab:medical_isotopes_with_number_density}
\end{table*}

In our survey of practical transmutation pathways, we found other radioistopes that cannot be made in as high quantity as those above because of insufficient feedstock isotopic abundance. Two particularly salient examples are the pathways $\ce{^192Pt(n,p)^192Pt}$ and $\ce{^186Os(n,p)^186Re}$. Producing $\ce{^111In}$ via $\quad \ce{^112_50Sn (n,d) ^111_49In}$ is also attractive, although we neglect it in this analysis due to the very high yield from $\ce{^112_50Sn (n, \mathrm{2n}) ^111_50Sn ->[{\beta{}^+}][{35\text{ min}}] ^111_49In}$.

As a general comparison of fission-product transmutation and fusion neutron-driven transmutation, we plot fission and fusion isotope yields in \Cref{fig:transmutation_space}. In \Cref{fig:transmutation_space}(a) we plot the relative yields from $\ce{^235U}$ fission versus neutron and proton number. We also highlight in red squares the radioisotopes produced in this work by fusion neutron transmutation. While there are a few radioisotopes that can be produced by fission - such as $\ce{^99Mo}$ and $\ce{^131I}$ - fusion driven transmutation can produce a much wider range of isotopes than fission. In \Cref{fig:transmutation_space}(b) we plot the relative fusion yield for all target isotopes produced by fusion neutron-driven transmutation on a monoisotopic feedstock. The relative fusion yield is defined as the number of target isotope nuclei produced per unit time. The yield is calculated using a Bateman equation \cite{bateman1910solution,cetnar2006general} solver. Only stable, abundant starting isotopes are considered. We also only consider reaction pathways that result in isotopes with different proton number to the feedstock, which allows chemical separation. The plots in \Cref{fig:transmutation_space} demonstrate the wide range of isotopes that can be produced with feedstock transmuted with fusion neutrons in addition to those considered here, potentially enabling new medical isotopes to be used. 

\section{Discussion} \label{sec:discussion}

This work shows that a megawatt-scale D–T neutron source could supply global demand for any single major medical radioisotope and enable production of emerging therapeutic and diagnostic nuclides. While the timeline for the deployment of such sources remains uncertain, rapid progress in compact fusion technology suggests that operational systems could become available within the next five years, and potentially sooner under accelerated development pathways. Further engagement between the medical radioisotope and the fusion energy communities will help motivate the development of these neutron sources in the near term. Fusion-neutron-driven radioisotope production quantities are summarized in \Cref{tab:medical_isotopes_with_number_density}.

Fusion-driven isotope production offers several advantages over conventional methods. A single neutron source can be reconfigured to produce a wide range of isotopes without reliance on fissile materials, therefore mitigating proliferation concerns and decoupling isotope supply chains from nuclear-fission infrastructure. This approach also enables on-demand, HSA production with simplified waste management and minimal long-lived byproducts.

Aside from requiring substantially smaller neutron rate (around 1000 times lower) and physical scale from fusion power plants, D-T fusion radioisotope producers also have relaxed fuel cycle constraints relative to larger systems.  Supplying every medical radioisotope market might require only around 10-20MW$_\mrm{th}$ of fusion power, or about 0.5-1kg/yr of tritium consumption.  Even in the absence of tritium breeding in the transmuter itself, this need could in principle be supplied by the CANDU reactor fleet, which produces about 3kg/yr of tritium \cite{kovari2017tritium}. Steady-state, high-flux D–T neutron sources with powers on the order of a megawatt are not yet available (as of 2025), but their development---distinct from power-plant-oriented fusion devices---is a recognized near-term priority \cite{abdou1995volumetric,kuteev2010intense,krolas2021ifmif,giannini2024conceptual,vigano2025multidisciplinary}.

Finally, although our analysis has focused on D–T neutron-driven transmutation, deuterium–deuterium (D–D) fusion neutron sources can also produce significant quantities of HSA radioisotopes on low-threshold targets such as $\ce{^{32}S(n,p)^{32}P}$, $\ce{^{47}Ti(n,p)^{47}Sc}$, $\ce{^{64}Zn(n,p)^{64}Cu}$, and $\ce{^{106}Cd(n,p)^{103}Pd}$; thus, even D–D fusion machines \cite{Hendel01101990,swanson2025scoping} are capable of producing several useful HSA radioisotopes.

Much remains to be done, particularly in the development of feedstock management and product extraction schemes, which will likely need to be different for each feedstock/product pair. Nonetheless, the path toward practical, fusion-based radioisotope production now appears increasingly achievable.

\section{Acknowledgments}

We are grateful for discussions with R. Delaporte-Mathurin, A. Diallo, and R. Radel.

\section{Availability of data and material}

The data used in this study will be made available on reasonable request.

\bibliography{Master_EverythingPlasmaBib} %

\begin{thebibliography}{101}%
\makeatletter
\providecommand \@ifxundefined [1]{%
 \@ifx{#1\undefined}
}%
\providecommand \@ifnum [1]{%
 \ifnum #1\expandafter \@firstoftwo
 \else \expandafter \@secondoftwo
 \fi
}%
\providecommand \@ifx [1]{%
 \ifx #1\expandafter \@firstoftwo
 \else \expandafter \@secondoftwo
 \fi
}%
\providecommand \natexlab [1]{#1}%
\providecommand \enquote  [1]{``#1''}%
\providecommand \bibnamefont  [1]{#1}%
\providecommand \bibfnamefont [1]{#1}%
\providecommand \citenamefont [1]{#1}%
\providecommand \href@noop [0]{\@secondoftwo}%
\providecommand \href [0]{\begingroup \@sanitize@url \@href}%
\providecommand \@href[1]{\@@startlink{#1}\@@href}%
\providecommand \@@href[1]{\endgroup#1\@@endlink}%
\providecommand \@sanitize@url [0]{\catcode `\\12\catcode `\$12\catcode
  `\&12\catcode `\#12\catcode `\^12\catcode `\_12\catcode `\%12\relax}%
\providecommand \@@startlink[1]{}%
\providecommand \@@endlink[0]{}%
\providecommand \url  [0]{\begingroup\@sanitize@url \@url }%
\providecommand \@url [1]{\endgroup\@href {#1}{\urlprefix }}%
\providecommand \urlprefix  [0]{URL }%
\providecommand \Eprint [0]{\href }%
\providecommand \doibase [0]{https://doi.org/}%
\providecommand \selectlanguage [0]{\@gobble}%
\providecommand \bibinfo  [0]{\@secondoftwo}%
\providecommand \bibfield  [0]{\@secondoftwo}%
\providecommand \translation [1]{[#1]}%
\providecommand \BibitemOpen [0]{}%
\providecommand \bibitemStop [0]{}%
\providecommand \bibitemNoStop [0]{.\EOS\space}%
\providecommand \EOS [0]{\spacefactor3000\relax}%
\providecommand \BibitemShut  [1]{\csname bibitem#1\endcsname}%
\let\auto@bib@innerbib\@empty
\bibitem [{\citenamefont {Weiner}\ and\ \citenamefont
  {Thakur}(1995)}]{weiner1995radionuclides}%
  \BibitemOpen
  \bibfield  {author} {\bibinfo {author} {\bibfnamefont {R.}~\bibnamefont
  {Weiner}}\ and\ \bibinfo {author} {\bibfnamefont {M.}~\bibnamefont
  {Thakur}},\ }\bibfield  {title} {\bibinfo {title} {Radionuclides:
  applications in diagnostic and therapeutic nuclear medicine},\ }\href@noop {}
  {\bibfield  {journal} {\bibinfo  {journal} {Radiochimica Acta}\ }\textbf
  {\bibinfo {volume} {70}},\ \bibinfo {pages} {273} (\bibinfo {year}
  {1995})}\BibitemShut {NoStop}%
\bibitem [{\citenamefont {Yeong}\ \emph {et~al.}(2014)\citenamefont {Yeong},
  \citenamefont {Cheng},\ and\ \citenamefont {Ng}}]{yeong2014therapeutic}%
  \BibitemOpen
  \bibfield  {author} {\bibinfo {author} {\bibfnamefont {C.-H.}\ \bibnamefont
  {Yeong}}, \bibinfo {author} {\bibfnamefont {M.-h.}\ \bibnamefont {Cheng}},\
  and\ \bibinfo {author} {\bibfnamefont {K.-H.}\ \bibnamefont {Ng}},\
  }\bibfield  {title} {\bibinfo {title} {Therapeutic radionuclides in nuclear
  medicine: current and future prospects},\ }\href@noop {} {\bibfield
  {journal} {\bibinfo  {journal} {Journal of Zhejiang University Science B}\
  }\textbf {\bibinfo {volume} {15}},\ \bibinfo {pages} {845} (\bibinfo {year}
  {2014})}\BibitemShut {NoStop}%
\bibitem [{\citenamefont {of~the
  European~Union)}(2009)}]{EC2009MedicalRadioisotopes}%
  \BibitemOpen
  \bibfield  {author} {\bibinfo {author} {\bibfnamefont {E.~C.~S.}\
  \bibnamefont {of~the European~Union)}},\ }\href
  {https://ec.europa.eu/health/ph_threats/radioisotopes/docs/radioisotopes_report_en.pdf}
  {\emph {\bibinfo {title} {Medical Radioisotopes --- Report of the Ad hoc
  Interservice Group of the European Commission on Sufficiency in Supply of
  Radioisotopes for Medical Use}}},\ \bibinfo {type} {Tech. Rep.}\ (\bibinfo
  {institution} {European Commission},\ \bibinfo {year} {2009})\ \bibinfo
  {note} {accessed: 2025-11-22}\BibitemShut {NoStop}%
\bibitem [{\citenamefont {Association}(2025)}]{WNA2025RadioIsotopes}%
  \BibitemOpen
  \bibfield  {author} {\bibinfo {author} {\bibfnamefont {W.~N.}\ \bibnamefont
  {Association}},\ }\href@noop {} {\bibinfo {title} {Radioisotopes in
  medicine}},\ \bibinfo {howpublished}
  {\url{https://world-nuclear.org/information-library/non-power-nuclear-applications/radioisotopes-research/radioisotopes-in-medicine}}
  (\bibinfo {year} {2025}),\ \bibinfo {note} {accessed: 2025-11-22}\BibitemShut
  {NoStop}%
\bibitem [{OEC(2011)}]{OECDNEA2011_Supply}%
  \BibitemOpen
  \href
  {https://www.oecd.org/content/dam/oecd/en/publications/reports/2011/06/the-supply-of-medical-radioisotopes_g1g37529/9789264991644-en.pdf}
  {\emph {\bibinfo {title} {The Supply of Medical Radioisotopes}}},\ \bibinfo
  {type} {Tech. Rep.}\ (\bibinfo  {institution} {OECD Nuclear Energy Agency},\
  \bibinfo {year} {2011})\BibitemShut {NoStop}%
\bibitem [{OEC(2010)}]{OECDNEA2010_Econ}%
  \BibitemOpen
  \href {https://www.oecd-nea.org/med-radio/reports/MO-99.pdf} {\emph {\bibinfo
  {title} {An Economic Study of the Molybdenum-99 Supply Chain}}},\ \bibinfo
  {type} {Tech. Rep.}\ (\bibinfo  {institution} {OECD Nuclear Energy Agency},\
  \bibinfo {year} {2010})\BibitemShut {NoStop}%
\bibitem [{NAS(2016)}]{NASEM_Mo99_2016}%
  \BibitemOpen
  \href {https://www.ncbi.nlm.nih.gov/books/NBK396159/} {\emph {\bibinfo
  {title} {Molybdenum-99 for Medical Imaging}}}\ (\bibinfo  {publisher}
  {National Academies Press},\ \bibinfo {year} {2016})\BibitemShut {NoStop}%
\bibitem [{\citenamefont {{SNMMI}}(2024)}]{SNMMI2024_Mo99_Tc99m}%
  \BibitemOpen
  \bibfield  {author} {\bibinfo {author} {\bibnamefont {{SNMMI}}},\ }\href
  {https://snmmi.org/Web/News/Articles/Imminent-Mo-99-Tc-99m-Shortage-Due-to-European-Reactor-Restart-Delay}
  {\bibinfo {title} {Imminent mo-99/tc-99m shortage due to european reactor
  restart delay}} (\bibinfo {year} {2024}),\ \bibinfo {note} {accessed:
  2025-10-24}\BibitemShut {NoStop}%
\bibitem [{IAE(2011)}]{IAEA_TRS473_2011}%
  \BibitemOpen
  \href {https://www-pub.iaea.org/MTCD/Publications/PDF/trs473_web.pdf} {\emph
  {\bibinfo {title} {Nuclear Data for the Production of Therapeutic
  Radionuclides}}},\ \bibinfo {type} {Tech. Rep.}\ \bibinfo {number}
  {IAEA-TRS-473}\ (\bibinfo  {institution} {IAEA},\ \bibinfo {year}
  {2011})\BibitemShut {NoStop}%
\bibitem [{\citenamefont {Habs}\ and\ \citenamefont
  {K{\"o}ster}(2011)}]{habs2011production}%
  \BibitemOpen
  \bibfield  {author} {\bibinfo {author} {\bibfnamefont {D.}~\bibnamefont
  {Habs}}\ and\ \bibinfo {author} {\bibfnamefont {U.}~\bibnamefont
  {K{\"o}ster}},\ }\bibfield  {title} {\bibinfo {title} {Production of medical
  radioisotopes with high specific activity in photonuclear reactions with
  $\gamma$-beams of high intensity and large brilliance},\ }\href@noop {}
  {\bibfield  {journal} {\bibinfo  {journal} {Applied Physics B}\ }\textbf
  {\bibinfo {volume} {103}},\ \bibinfo {pages} {501} (\bibinfo {year}
  {2011})}\BibitemShut {NoStop}%
\bibitem [{\citenamefont {Hansen}\ and\ \citenamefont
  {Bender}(2022)}]{hansen2022advancement}%
  \BibitemOpen
  \bibfield  {author} {\bibinfo {author} {\bibfnamefont {S.~B.}\ \bibnamefont
  {Hansen}}\ and\ \bibinfo {author} {\bibfnamefont {D.}~\bibnamefont
  {Bender}},\ }\bibfield  {title} {\bibinfo {title} {Advancement in production
  of radiotracers},\ }in\ \href@noop {} {\emph {\bibinfo {booktitle} {Seminars
  in nuclear medicine}}},\ Vol.~\bibinfo {volume} {52}\ (\bibinfo
  {organization} {Elsevier},\ \bibinfo {year} {2022})\ pp.\ \bibinfo {pages}
  {266--275}\BibitemShut {NoStop}%
\bibitem [{\citenamefont {Strachan}\ \emph {et~al.}(1994)\citenamefont
  {Strachan} \emph {et~al.}}]{Strachan1994short}%
  \BibitemOpen
  \bibfield  {author} {\bibinfo {author} {\bibfnamefont {J.~D.}\ \bibnamefont
  {Strachan}} \emph {et~al.},\ }\bibfield  {title} {\bibinfo {title} {Fusion
  power production from tftr plasmas fueled with deuterium and tritium},\
  }\bibfield  {journal} {\bibinfo  {journal} {Physical Review Letters}\
  }\textbf {\bibinfo {volume} {72}},\ \href
  {https://doi.org/10.1103/PhysRevLett.72.3526} {10.1103/PhysRevLett.72.3526}
  (\bibinfo {year} {1994})\BibitemShut {NoStop}%
\bibitem [{\citenamefont {Keilhacker}\ \emph {et~al.}(1999)\citenamefont
  {Keilhacker}, \citenamefont {Gibson}, \citenamefont {Gormezano},
  \citenamefont {Lomas}, \citenamefont {Thomas}, \citenamefont {Watkins},
  \citenamefont {Andrew}, \citenamefont {Balet}, \citenamefont {Borba},
  \citenamefont {Challis}, \citenamefont {Coffey}, \citenamefont {Cottrell},
  \citenamefont {Esch}, \citenamefont {Deliyanakis}, \citenamefont {Fasoli},
  \citenamefont {Gowers}, \citenamefont {Guo}, \citenamefont {Huysmans},
  \citenamefont {Jones}, \citenamefont {Kerner}, \citenamefont {K{\"{o,}}nig},
  \citenamefont {Loughlin}, \citenamefont {Maas}, \citenamefont {Marcus},
  \citenamefont {Nave}, \citenamefont {Rimini}, \citenamefont {Sadler},
  \citenamefont {Sharapov}, \citenamefont {Sips}, \citenamefont {Smeulders},
  \citenamefont {S{\"{o}}ldner}, \citenamefont {Taroni}, \citenamefont
  {Tubbing}, \citenamefont {von Hellermann}, \citenamefont {Ward},\ and\
  \citenamefont {Team}}]{Keilhacker1999}%
  \BibitemOpen
  \bibfield  {author} {\bibinfo {author} {\bibfnamefont {M.}~\bibnamefont
  {Keilhacker}}, \bibinfo {author} {\bibfnamefont {A.}~\bibnamefont {Gibson}},
  \bibinfo {author} {\bibfnamefont {C.}~\bibnamefont {Gormezano}}, \bibinfo
  {author} {\bibfnamefont {P.~J.}\ \bibnamefont {Lomas}}, \bibinfo {author}
  {\bibfnamefont {P.~R.}\ \bibnamefont {Thomas}}, \bibinfo {author}
  {\bibfnamefont {M.~L.}\ \bibnamefont {Watkins}}, \bibinfo {author}
  {\bibfnamefont {P.}~\bibnamefont {Andrew}}, \bibinfo {author} {\bibfnamefont
  {B.}~\bibnamefont {Balet}}, \bibinfo {author} {\bibfnamefont
  {D.}~\bibnamefont {Borba}}, \bibinfo {author} {\bibfnamefont {C.~D.}\
  \bibnamefont {Challis}}, \bibinfo {author} {\bibfnamefont {I.}~\bibnamefont
  {Coffey}}, \bibinfo {author} {\bibfnamefont {G.~A.}\ \bibnamefont
  {Cottrell}}, \bibinfo {author} {\bibfnamefont {H.~P. L.~D.}\ \bibnamefont
  {Esch}}, \bibinfo {author} {\bibfnamefont {N.}~\bibnamefont {Deliyanakis}},
  \bibinfo {author} {\bibfnamefont {A.}~\bibnamefont {Fasoli}}, \bibinfo
  {author} {\bibfnamefont {C.~W.}\ \bibnamefont {Gowers}}, \bibinfo {author}
  {\bibfnamefont {H.~Y.}\ \bibnamefont {Guo}}, \bibinfo {author} {\bibfnamefont
  {G.~T.~A.}\ \bibnamefont {Huysmans}}, \bibinfo {author} {\bibfnamefont
  {T.~T.~C.}\ \bibnamefont {Jones}}, \bibinfo {author} {\bibfnamefont
  {W.}~\bibnamefont {Kerner}}, \bibinfo {author} {\bibfnamefont
  {R.}~\bibnamefont {K{\"{o,}}nig}}, \bibinfo {author} {\bibfnamefont
  {M.}~\bibnamefont {Loughlin}}, \bibinfo {author} {\bibfnamefont
  {A.}~\bibnamefont {Maas}}, \bibinfo {author} {\bibfnamefont {F.}~\bibnamefont
  {Marcus}}, \bibinfo {author} {\bibfnamefont {M.}~\bibnamefont {Nave}},
  \bibinfo {author} {\bibfnamefont {F.}~\bibnamefont {Rimini}}, \bibinfo
  {author} {\bibfnamefont {G.}~\bibnamefont {Sadler}}, \bibinfo {author}
  {\bibfnamefont {S.}~\bibnamefont {Sharapov}}, \bibinfo {author}
  {\bibfnamefont {G.}~\bibnamefont {Sips}}, \bibinfo {author} {\bibfnamefont
  {P.}~\bibnamefont {Smeulders}}, \bibinfo {author} {\bibfnamefont
  {F.}~\bibnamefont {S{\"{o}}ldner}}, \bibinfo {author} {\bibfnamefont
  {A.}~\bibnamefont {Taroni}}, \bibinfo {author} {\bibfnamefont
  {B.}~\bibnamefont {Tubbing}}, \bibinfo {author} {\bibfnamefont
  {M.}~\bibnamefont {von Hellermann}}, \bibinfo {author} {\bibfnamefont
  {D.}~\bibnamefont {Ward}},\ and\ \bibinfo {author} {\bibfnamefont
  {J.}~\bibnamefont {Team}},\ }\bibfield  {title} {\bibinfo {title} {{High
  fusion performance from deuterium-tritium plasmas in JET}},\ }\href@noop {}
  {\bibfield  {journal} {\bibinfo  {journal} {Nuclear Fusion}\ }\textbf
  {\bibinfo {volume} {39}},\ \bibinfo {pages} {209} (\bibinfo {year}
  {1999})}\BibitemShut {NoStop}%
\bibitem [{\citenamefont {Wurzel}\ and\ \citenamefont
  {Hsu}(2022)}]{Wurzel2022}%
  \BibitemOpen
  \bibfield  {author} {\bibinfo {author} {\bibfnamefont {S.~E.}\ \bibnamefont
  {Wurzel}}\ and\ \bibinfo {author} {\bibfnamefont {S.~C.}\ \bibnamefont
  {Hsu}},\ }\bibfield  {title} {\bibinfo {title} {Progress toward fusion energy
  breakeven and gain as measured against the lawson criterion},\ }\bibfield
  {journal} {\bibinfo  {journal} {Physics of Plasmas}\ }\textbf {\bibinfo
  {volume} {29}},\ \href {https://doi.org/10.1063/5.0083990}
  {10.1063/5.0083990} (\bibinfo {year} {2022})\BibitemShut {NoStop}%
\bibitem [{\citenamefont {Engholm}\ \emph {et~al.}(1986)\citenamefont
  {Engholm}, \citenamefont {Cheng},\ and\ \citenamefont
  {Schultz}}]{engholm1986radioisotope}%
  \BibitemOpen
  \bibfield  {author} {\bibinfo {author} {\bibfnamefont {B.~A.}\ \bibnamefont
  {Engholm}}, \bibinfo {author} {\bibfnamefont {E.~T.}\ \bibnamefont {Cheng}},\
  and\ \bibinfo {author} {\bibfnamefont {K.~R.}\ \bibnamefont {Schultz}},\
  }\bibfield  {title} {\bibinfo {title} {Radioisotope production in fusion
  reactors},\ }\href@noop {} {\bibfield  {journal} {\bibinfo  {journal} {Fusion
  technology}\ }\textbf {\bibinfo {volume} {10}},\ \bibinfo {pages} {1290}
  (\bibinfo {year} {1986})}\BibitemShut {NoStop}%
\bibitem [{\citenamefont {Bourque}\ \emph {et~al.}(1988)\citenamefont
  {Bourque}, \citenamefont {Schultz},\ and\ \citenamefont
  {Staff}}]{Bourque1988FAME}%
  \BibitemOpen
  \bibfield  {author} {\bibinfo {author} {\bibfnamefont {R.}~\bibnamefont
  {Bourque}}, \bibinfo {author} {\bibfnamefont {K.}~\bibnamefont {Schultz}},\
  and\ \bibinfo {author} {\bibfnamefont {P.}~\bibnamefont {Staff}},\
  }\href@noop {} {\emph {\bibinfo {title} {Fusion Applications and Market
  Evaluation (FAME) Study}}},\ \bibinfo {type} {Technical Report}\ \bibinfo
  {number} {GA\mbox{-}A18658 / UCRL\mbox{-}21073 / UC\mbox{-}420 /
  UC\mbox{-}424 / UC\mbox{-}712}\ (\bibinfo  {institution} {GA Technologies,
  Inc.\ (General Atomics)},\ \bibinfo {address} {San Diego, CA},\ \bibinfo
  {year} {1988})\ \bibinfo {note} {prepared under Subcontract 8236305 for
  Lawrence Livermore National Laboratory; DTIC accession AD\mbox{-}A243
  768}\BibitemShut {NoStop}%
\bibitem [{\citenamefont {Ridikas}\ \emph {et~al.}(2006)\citenamefont
  {Ridikas}, \citenamefont {Plukiene}, \citenamefont {Plukis},\ and\
  \citenamefont {Cheng}}]{Ridikas2006_HybridWaste}%
  \BibitemOpen
  \bibfield  {author} {\bibinfo {author} {\bibfnamefont {D.}~\bibnamefont
  {Ridikas}}, \bibinfo {author} {\bibfnamefont {R.}~\bibnamefont {Plukiene}},
  \bibinfo {author} {\bibfnamefont {A.}~\bibnamefont {Plukis}},\ and\ \bibinfo
  {author} {\bibfnamefont {E.~T.}\ \bibnamefont {Cheng}},\ }\bibfield  {title}
  {\bibinfo {title} {Fusion--fission hybrid system for nuclear waste
  transmutation (i): Characterization of the system and burn-up calculations},\
  }\href {https://doi.org/10.1016/j.pnucene.2005.09.004} {\bibfield  {journal}
  {\bibinfo  {journal} {Progress in Nuclear Energy}\ }\textbf {\bibinfo
  {volume} {48}},\ \bibinfo {pages} {235} (\bibinfo {year} {2006})}\BibitemShut
  {NoStop}%
\bibitem [{\citenamefont {Leung}\ \emph {et~al.}(2018)\citenamefont {Leung},
  \citenamefont {Leung},\ and\ \citenamefont {Melville}}]{Leung2018_CompactNG}%
  \BibitemOpen
  \bibfield  {author} {\bibinfo {author} {\bibfnamefont {K.~N.}\ \bibnamefont
  {Leung}}, \bibinfo {author} {\bibfnamefont {J.~K.}\ \bibnamefont {Leung}},\
  and\ \bibinfo {author} {\bibfnamefont {G.}~\bibnamefont {Melville}},\
  }\bibfield  {title} {\bibinfo {title} {Feasibility study on medical isotope
  production using a compact neutron generator},\ }\href
  {https://doi.org/10.1016/j.apradiso.2018.02.026} {\bibfield  {journal}
  {\bibinfo  {journal} {Applied Radiation and Isotopes}\ }\textbf {\bibinfo
  {volume} {137}},\ \bibinfo {pages} {23} (\bibinfo {year} {2018})}\BibitemShut
  {NoStop}%
\bibitem [{\citenamefont {Pietropaolo}\ \emph {et~al.}(2021)\citenamefont
  {Pietropaolo}, \citenamefont {Contessa}, \citenamefont {Farini},
  \citenamefont {Fonnesu}, \citenamefont {Marinari}, \citenamefont {Moro},
  \citenamefont {Rizzo}, \citenamefont {Scaglione}, \citenamefont {Terranova},
  \citenamefont {Utili} \emph {et~al.}}]{pietropaolo2021sorgentina}%
  \BibitemOpen
  \bibfield  {author} {\bibinfo {author} {\bibfnamefont {A.}~\bibnamefont
  {Pietropaolo}}, \bibinfo {author} {\bibfnamefont {G.~M.}\ \bibnamefont
  {Contessa}}, \bibinfo {author} {\bibfnamefont {M.}~\bibnamefont {Farini}},
  \bibinfo {author} {\bibfnamefont {N.}~\bibnamefont {Fonnesu}}, \bibinfo
  {author} {\bibfnamefont {R.}~\bibnamefont {Marinari}}, \bibinfo {author}
  {\bibfnamefont {F.}~\bibnamefont {Moro}}, \bibinfo {author} {\bibfnamefont
  {A.}~\bibnamefont {Rizzo}}, \bibinfo {author} {\bibfnamefont
  {S.}~\bibnamefont {Scaglione}}, \bibinfo {author} {\bibfnamefont
  {N.}~\bibnamefont {Terranova}}, \bibinfo {author} {\bibfnamefont
  {M.}~\bibnamefont {Utili}}, \emph {et~al.},\ }\bibfield  {title} {\bibinfo
  {title} {Sorgentina-rf project: Fusion neutrons for 99mo medical
  radioisotope},\ }\href@noop {} {\bibfield  {journal} {\bibinfo  {journal}
  {Eur. Phys. J. Plus}\ }\textbf {\bibinfo {volume} {136}},\ \bibinfo {pages}
  {1140} (\bibinfo {year} {2021})}\BibitemShut {NoStop}%
\bibitem [{\citenamefont {Honney}(2023)}]{Honney2023FusionNeutrons}%
  \BibitemOpen
  \bibfield  {author} {\bibinfo {author} {\bibfnamefont {T.}~\bibnamefont
  {Honney}},\ }\bibfield  {title} {\bibinfo {title} {New value from fusion
  neutrons},\ }\href
  {https://www.neimagazine.com/analysis/new-value-from-fusion-neutrons-10924344/}
  {\bibfield  {journal} {\bibinfo  {journal} {Nuclear Engineering
  International}\ } (\bibinfo {year} {2023})},\ \bibinfo {note} {interview with
  Greg Piefer (SHINE Technologies)}\BibitemShut {NoStop}%
\bibitem [{\citenamefont {Pereslavtsev}\ \emph {et~al.}(2024)\citenamefont
  {Pereslavtsev}, \citenamefont {Bachmann}, \citenamefont {Elbez-Uzan},\ and\
  \citenamefont {Park}}]{pereslavtsev2024potential}%
  \BibitemOpen
  \bibfield  {author} {\bibinfo {author} {\bibfnamefont {P.}~\bibnamefont
  {Pereslavtsev}}, \bibinfo {author} {\bibfnamefont {C.}~\bibnamefont
  {Bachmann}}, \bibinfo {author} {\bibfnamefont {J.}~\bibnamefont
  {Elbez-Uzan}},\ and\ \bibinfo {author} {\bibfnamefont {J.~H.}\ \bibnamefont
  {Park}},\ }\bibfield  {title} {\bibinfo {title} {Potential of radioactive
  isotopes production in demo for commercial use},\ }\href@noop {} {\bibfield
  {journal} {\bibinfo  {journal} {Applied Sciences}\ }\textbf {\bibinfo
  {volume} {14}},\ \bibinfo {pages} {442} (\bibinfo {year} {2024})}\BibitemShut
  {NoStop}%
\bibitem [{\citenamefont {Ponsard}(2014)}]{ponsard2014production}%
  \BibitemOpen
  \bibfield  {author} {\bibinfo {author} {\bibfnamefont {B.}~\bibnamefont
  {Ponsard}},\ }\bibfield  {title} {\bibinfo {title} {Production of lu-177 in
  the br2 high-flux reactor},\ }\href@noop {} {\bibfield  {journal} {\bibinfo
  {journal} {Nuclear Medicine and Biology}\ }\textbf {\bibinfo {volume} {7}},\
  \bibinfo {pages} {648} (\bibinfo {year} {2014})}\BibitemShut {NoStop}%
\bibitem [{\citenamefont {Lederer-Woods}\ \emph {et~al.}(2024)\citenamefont
  {Lederer-Woods}, \citenamefont {Sosnin}, \citenamefont {Woods}, \citenamefont
  {n\_TOF Collaboration} \emph {et~al.}}]{lederer2024measurement}%
  \BibitemOpen
  \bibfield  {author} {\bibinfo {author} {\bibfnamefont {C.}~\bibnamefont
  {Lederer-Woods}}, \bibinfo {author} {\bibfnamefont {N.}~\bibnamefont
  {Sosnin}}, \bibinfo {author} {\bibfnamefont {P.}~\bibnamefont {Woods}},
  \bibinfo {author} {\bibnamefont {n\_TOF Collaboration}}, \emph {et~al.},\
  }\bibfield  {title} {\bibinfo {title} {Measurement of the yb 176 (n,
  $\gamma$) cross section at the n\_tof facility at cern},\ }\href@noop {}
  {\bibfield  {journal} {\bibinfo  {journal} {Physical Review C}\ }\textbf
  {\bibinfo {volume} {110}},\ \bibinfo {pages} {1} (\bibinfo {year}
  {2024})}\BibitemShut {NoStop}%
\bibitem [{\citenamefont {Li}\ and\ \citenamefont
  {Zheng}(2023)}]{li2023feasibility}%
  \BibitemOpen
  \bibfield  {author} {\bibinfo {author} {\bibfnamefont {J.}~\bibnamefont
  {Li}}\ and\ \bibinfo {author} {\bibfnamefont {S.}~\bibnamefont {Zheng}},\
  }\bibfield  {title} {\bibinfo {title} {Feasibility study to byproduce medical
  radioisotopes in a fusion reactor},\ }\href@noop {} {\bibfield  {journal}
  {\bibinfo  {journal} {Molecules}\ }\textbf {\bibinfo {volume} {28}},\
  \bibinfo {pages} {2040} (\bibinfo {year} {2023})}\BibitemShut {NoStop}%
\bibitem [{\citenamefont {Evitts}\ \emph {et~al.}()\citenamefont {Evitts},
  \citenamefont {Miller}, \citenamefont {{Da Pieve}}, \citenamefont {Turner},\
  and\ \citenamefont {Borini}}]{evitts2025theoretical}%
  \BibitemOpen
  \bibfield  {author} {\bibinfo {author} {\bibfnamefont {L.~J.}\ \bibnamefont
  {Evitts}}, \bibinfo {author} {\bibfnamefont {P.~W.}\ \bibnamefont {Miller}},
  \bibinfo {author} {\bibfnamefont {C.}~\bibnamefont {{Da Pieve}}}, \bibinfo
  {author} {\bibfnamefont {A.}~\bibnamefont {Turner}},\ and\ \bibinfo {author}
  {\bibfnamefont {S.}~\bibnamefont {Borini}},\ }\bibfield  {title} {\bibinfo
  {title} {Theoretical novel medical isotope production with deuterium-tritium
  fusion technology},\ }\href
  {https://doi.org/https://doi.org/10.1016/j.apradiso.2025.112163} {\bibfield
  {journal} {\bibinfo  {journal} {Applied Radiation and Isotopes}\ }\textbf
  {\bibinfo {volume} {226}},\ \bibinfo {pages} {112163}}\BibitemShut {NoStop}%
\bibitem [{\citenamefont {Rutkowski}\ \emph {et~al.}(2025)\citenamefont
  {Rutkowski}, \citenamefont {Harter},\ and\ \citenamefont
  {Parisi}}]{Rutkowski2025}%
  \BibitemOpen
  \bibfield  {author} {\bibinfo {author} {\bibfnamefont {A.}~\bibnamefont
  {Rutkowski}}, \bibinfo {author} {\bibfnamefont {J.}~\bibnamefont {Harter}},\
  and\ \bibinfo {author} {\bibfnamefont {J.}~\bibnamefont {Parisi}},\ }\href
  {https://arxiv.org/abs/2507.13461} {\bibinfo {title} {Scalable chrysopoeia
  via $(n, 2n)$ reactions driven by deuterium-tritium fusion neutrons}}
  (\bibinfo {year} {2025}),\ \Eprint {https://arxiv.org/abs/2507.13461}
  {arXiv:2507.13461 [physics.plasm-ph]} \BibitemShut {NoStop}%
\bibitem [{\citenamefont {Nichols}\ and\ \citenamefont
  {Binford}(1971)}]{nichols1971status}%
  \BibitemOpen
  \bibfield  {author} {\bibinfo {author} {\bibfnamefont {J.}~\bibnamefont
  {Nichols}}\ and\ \bibinfo {author} {\bibfnamefont {F.}~\bibnamefont
  {Binford}},\ }\href@noop {} {\emph {\bibinfo {title} {Status OF Noble Gas
  Removal And Disposal.}}},\ \bibinfo {type} {Tech. Rep.}\ (\bibinfo
  {institution} {Oak Ridge National Lab.(ORNL), Oak Ridge, TN (United
  States)},\ \bibinfo {year} {1971})\BibitemShut {NoStop}%
\bibitem [{\citenamefont {Maroni}\ \emph {et~al.}(1975)\citenamefont {Maroni},
  \citenamefont {Wolson},\ and\ \citenamefont {Staahl}}]{maroni1975some}%
  \BibitemOpen
  \bibfield  {author} {\bibinfo {author} {\bibfnamefont {V.~A.}\ \bibnamefont
  {Maroni}}, \bibinfo {author} {\bibfnamefont {R.~D.}\ \bibnamefont {Wolson}},\
  and\ \bibinfo {author} {\bibfnamefont {G.~E.}\ \bibnamefont {Staahl}},\
  }\bibfield  {title} {\bibinfo {title} {Some preliminary considerations of a
  molten-salt extraction process to remove tritium from liquid lithium fusion
  reactor blankets},\ }\href@noop {} {\bibfield  {journal} {\bibinfo  {journal}
  {Nuclear Technology}\ }\textbf {\bibinfo {volume} {25}},\ \bibinfo {pages}
  {83} (\bibinfo {year} {1975})}\BibitemShut {NoStop}%
\bibitem [{\citenamefont {Horwitz}\ \emph {et~al.}(2005)\citenamefont
  {Horwitz}, \citenamefont {McAlister}, \citenamefont {Bond}, \citenamefont
  {Barrans},\ and\ \citenamefont {Williamson}}]{horwitz2005process}%
  \BibitemOpen
  \bibfield  {author} {\bibinfo {author} {\bibfnamefont {E.}~\bibnamefont
  {Horwitz}}, \bibinfo {author} {\bibfnamefont {D.}~\bibnamefont {McAlister}},
  \bibinfo {author} {\bibfnamefont {A.}~\bibnamefont {Bond}}, \bibinfo {author}
  {\bibfnamefont {R.}~\bibnamefont {Barrans}},\ and\ \bibinfo {author}
  {\bibfnamefont {J.}~\bibnamefont {Williamson}},\ }\bibfield  {title}
  {\bibinfo {title} {A process for the separation of 177lu from neutron
  irradiated 176yb targets},\ }\href@noop {} {\bibfield  {journal} {\bibinfo
  {journal} {Applied Radiation and Isotopes}\ }\textbf {\bibinfo {volume}
  {63}},\ \bibinfo {pages} {23} (\bibinfo {year} {2005})}\BibitemShut {NoStop}%
\bibitem [{\citenamefont {Chattopadhyay}\ and\ \citenamefont
  {Das}(2009)}]{chattopadhyay2009simple}%
  \BibitemOpen
  \bibfield  {author} {\bibinfo {author} {\bibfnamefont {S.}~\bibnamefont
  {Chattopadhyay}}\ and\ \bibinfo {author} {\bibfnamefont {S.~S.}\ \bibnamefont
  {Das}},\ }\bibfield  {title} {\bibinfo {title} {A simple and rapid technique
  for radiochemical separation of iodine radionuclides from irradiated
  tellurium using an activated charcoal column},\ }\href@noop {} {\bibfield
  {journal} {\bibinfo  {journal} {Applied Radiation and Isotopes}\ }\textbf
  {\bibinfo {volume} {67}},\ \bibinfo {pages} {1748} (\bibinfo {year}
  {2009})}\BibitemShut {NoStop}%
\bibitem [{\citenamefont {Demange}\ \emph {et~al.}(2016)\citenamefont
  {Demange}, \citenamefont {Antunes}, \citenamefont {Borisevich}, \citenamefont
  {Frances}, \citenamefont {Rapisarda}, \citenamefont {Santucci},\ and\
  \citenamefont {Utili}}]{demange2016tritium}%
  \BibitemOpen
  \bibfield  {author} {\bibinfo {author} {\bibfnamefont {D.}~\bibnamefont
  {Demange}}, \bibinfo {author} {\bibfnamefont {R.}~\bibnamefont {Antunes}},
  \bibinfo {author} {\bibfnamefont {O.}~\bibnamefont {Borisevich}}, \bibinfo
  {author} {\bibfnamefont {L.}~\bibnamefont {Frances}}, \bibinfo {author}
  {\bibfnamefont {D.}~\bibnamefont {Rapisarda}}, \bibinfo {author}
  {\bibfnamefont {A.}~\bibnamefont {Santucci}},\ and\ \bibinfo {author}
  {\bibfnamefont {M.}~\bibnamefont {Utili}},\ }\bibfield  {title} {\bibinfo
  {title} {Tritium extraction technologies and demo requirements},\ }\href@noop
  {} {\bibfield  {journal} {\bibinfo  {journal} {Fusion Engineering and
  Design}\ }\textbf {\bibinfo {volume} {109}},\ \bibinfo {pages} {912}
  (\bibinfo {year} {2016})}\BibitemShut {NoStop}%
\bibitem [{\citenamefont {Cristescu}\ and\ \citenamefont
  {Draghia}(2020)}]{cristescu2020developments}%
  \BibitemOpen
  \bibfield  {author} {\bibinfo {author} {\bibfnamefont {I.}~\bibnamefont
  {Cristescu}}\ and\ \bibinfo {author} {\bibfnamefont {M.}~\bibnamefont
  {Draghia}},\ }\bibfield  {title} {\bibinfo {title} {Developments on the
  tritium extraction and recovery system for hcpb},\ }\href@noop {} {\bibfield
  {journal} {\bibinfo  {journal} {Fusion Engineering and Design}\ }\textbf
  {\bibinfo {volume} {158}},\ \bibinfo {pages} {111558} (\bibinfo {year}
  {2020})}\BibitemShut {NoStop}%
\bibitem [{\citenamefont {Holiski}\ \emph {et~al.}(2024)\citenamefont
  {Holiski}, \citenamefont {Bender}, \citenamefont {Monte}, \citenamefont
  {Hennkens}, \citenamefont {Embree}, \citenamefont {Wang}, \citenamefont
  {Sjoden},\ and\ \citenamefont {Mastren}}]{holiski2024production}%
  \BibitemOpen
  \bibfield  {author} {\bibinfo {author} {\bibfnamefont {C.~K.}\ \bibnamefont
  {Holiski}}, \bibinfo {author} {\bibfnamefont {A.~A.}\ \bibnamefont {Bender}},
  \bibinfo {author} {\bibfnamefont {P.~F.}\ \bibnamefont {Monte}}, \bibinfo
  {author} {\bibfnamefont {H.~M.}\ \bibnamefont {Hennkens}}, \bibinfo {author}
  {\bibfnamefont {M.~F.}\ \bibnamefont {Embree}}, \bibinfo {author}
  {\bibfnamefont {M.-J.~V.}\ \bibnamefont {Wang}}, \bibinfo {author}
  {\bibfnamefont {G.~E.}\ \bibnamefont {Sjoden}},\ and\ \bibinfo {author}
  {\bibfnamefont {T.}~\bibnamefont {Mastren}},\ }\bibfield  {title} {\bibinfo
  {title} {The production and separation of 161tb with high specific activity
  at the university of utah},\ }\href@noop {} {\bibfield  {journal} {\bibinfo
  {journal} {Applied Radiation and Isotopes}\ }\textbf {\bibinfo {volume}
  {214}},\ \bibinfo {pages} {111530} (\bibinfo {year} {2024})}\BibitemShut
  {NoStop}%
\bibitem [{\citenamefont {Terrell}(1962)}]{terrell1962}%
  \BibitemOpen
  \bibfield  {author} {\bibinfo {author} {\bibfnamefont {J.}~\bibnamefont
  {Terrell}},\ }\bibfield  {title} {\bibinfo {title} {Neutron yields from
  individual fission fragments},\ }\href
  {https://doi.org/10.1103/PhysRev.127.880} {\bibfield  {journal} {\bibinfo
  {journal} {Phys. Rev.}\ }\textbf {\bibinfo {volume} {127}},\ \bibinfo {pages}
  {880} (\bibinfo {year} {1962})}\BibitemShut {NoStop}%
\bibitem [{\citenamefont {Brown}\ \emph {et~al.}(2018)\citenamefont {Brown},
  \citenamefont {Chadwick}, \citenamefont {Capote}, \citenamefont {Kahler},
  \citenamefont {Trkov}, \citenamefont {Herman}, \citenamefont {Sonzogni},
  \citenamefont {Danon}, \citenamefont {Carlson}, \citenamefont {Dunn},
  \citenamefont {Smith}, \citenamefont {Hale}, \citenamefont {Arbanas},
  \citenamefont {Arcilla}, \citenamefont {Bates}, \citenamefont {Beck},
  \citenamefont {Becker}, \citenamefont {Brown}, \citenamefont {Casperson},
  \citenamefont {Conlin}, \citenamefont {Cullen}, \citenamefont {Descalle},
  \citenamefont {Firestone}, \citenamefont {Gaines}, \citenamefont {Guber},
  \citenamefont {Hawari}, \citenamefont {Holmes}, \citenamefont {Johnson},
  \citenamefont {Kawano}, \citenamefont {Kiedrowski}, \citenamefont {Koning},
  \citenamefont {Kopecky}, \citenamefont {Leal}, \citenamefont {Lestone},
  \citenamefont {Lubitz}, \citenamefont {Dami{\'a}n}, \citenamefont {Mattoon},
  \citenamefont {McCutchan}, \citenamefont {Mughabghab}, \citenamefont
  {Navratil}, \citenamefont {Neudecker}, \citenamefont {Nobre}, \citenamefont
  {Noguere}, \citenamefont {Paris}, \citenamefont {Pigni}, \citenamefont
  {Plompen}, \citenamefont {Pritychenko}, \citenamefont {Pronyaev},
  \citenamefont {Roubtsov}, \citenamefont {Rochman}, \citenamefont {Romano},
  \citenamefont {Schillebeeckx}, \citenamefont {Simakov}, \citenamefont {Sin},
  \citenamefont {Sirakov}, \citenamefont {Sleaford}, \citenamefont {Sobes},
  \citenamefont {Soukhovitskii}, \citenamefont {Stetcu}, \citenamefont {Talou},
  \citenamefont {Thompson}, \citenamefont {van~der Marck}, \citenamefont
  {Welser-Sherrill}, \citenamefont {Wiarda}, \citenamefont {White},
  \citenamefont {Wormald}, \citenamefont {Wright}, \citenamefont {Zerkle},
  \citenamefont {\v{Z}erovnik},\ and\ \citenamefont {Zhu}}]{Brown20181}%
  \BibitemOpen
  \bibfield  {author} {\bibinfo {author} {\bibfnamefont {D.}~\bibnamefont
  {Brown}}, \bibinfo {author} {\bibfnamefont {M.}~\bibnamefont {Chadwick}},
  \bibinfo {author} {\bibfnamefont {R.}~\bibnamefont {Capote}}, \bibinfo
  {author} {\bibfnamefont {A.}~\bibnamefont {Kahler}}, \bibinfo {author}
  {\bibfnamefont {A.}~\bibnamefont {Trkov}}, \bibinfo {author} {\bibfnamefont
  {M.}~\bibnamefont {Herman}}, \bibinfo {author} {\bibfnamefont
  {A.}~\bibnamefont {Sonzogni}}, \bibinfo {author} {\bibfnamefont
  {Y.}~\bibnamefont {Danon}}, \bibinfo {author} {\bibfnamefont
  {A.}~\bibnamefont {Carlson}}, \bibinfo {author} {\bibfnamefont
  {M.}~\bibnamefont {Dunn}}, \bibinfo {author} {\bibfnamefont {D.}~\bibnamefont
  {Smith}}, \bibinfo {author} {\bibfnamefont {G.}~\bibnamefont {Hale}},
  \bibinfo {author} {\bibfnamefont {G.}~\bibnamefont {Arbanas}}, \bibinfo
  {author} {\bibfnamefont {R.}~\bibnamefont {Arcilla}}, \bibinfo {author}
  {\bibfnamefont {C.}~\bibnamefont {Bates}}, \bibinfo {author} {\bibfnamefont
  {B.}~\bibnamefont {Beck}}, \bibinfo {author} {\bibfnamefont {B.}~\bibnamefont
  {Becker}}, \bibinfo {author} {\bibfnamefont {F.}~\bibnamefont {Brown}},
  \bibinfo {author} {\bibfnamefont {R.}~\bibnamefont {Casperson}}, \bibinfo
  {author} {\bibfnamefont {J.}~\bibnamefont {Conlin}}, \bibinfo {author}
  {\bibfnamefont {D.}~\bibnamefont {Cullen}}, \bibinfo {author} {\bibfnamefont
  {M.-A.}\ \bibnamefont {Descalle}}, \bibinfo {author} {\bibfnamefont
  {R.}~\bibnamefont {Firestone}}, \bibinfo {author} {\bibfnamefont
  {T.}~\bibnamefont {Gaines}}, \bibinfo {author} {\bibfnamefont
  {K.}~\bibnamefont {Guber}}, \bibinfo {author} {\bibfnamefont
  {A.}~\bibnamefont {Hawari}}, \bibinfo {author} {\bibfnamefont
  {J.}~\bibnamefont {Holmes}}, \bibinfo {author} {\bibfnamefont
  {T.}~\bibnamefont {Johnson}}, \bibinfo {author} {\bibfnamefont
  {T.}~\bibnamefont {Kawano}}, \bibinfo {author} {\bibfnamefont
  {B.}~\bibnamefont {Kiedrowski}}, \bibinfo {author} {\bibfnamefont
  {A.}~\bibnamefont {Koning}}, \bibinfo {author} {\bibfnamefont
  {S.}~\bibnamefont {Kopecky}}, \bibinfo {author} {\bibfnamefont
  {L.}~\bibnamefont {Leal}}, \bibinfo {author} {\bibfnamefont {J.}~\bibnamefont
  {Lestone}}, \bibinfo {author} {\bibfnamefont {C.}~\bibnamefont {Lubitz}},
  \bibinfo {author} {\bibfnamefont {J.~M.}\ \bibnamefont {Dami{\'a}n}},
  \bibinfo {author} {\bibfnamefont {C.}~\bibnamefont {Mattoon}}, \bibinfo
  {author} {\bibfnamefont {E.}~\bibnamefont {McCutchan}}, \bibinfo {author}
  {\bibfnamefont {S.}~\bibnamefont {Mughabghab}}, \bibinfo {author}
  {\bibfnamefont {P.}~\bibnamefont {Navratil}}, \bibinfo {author}
  {\bibfnamefont {D.}~\bibnamefont {Neudecker}}, \bibinfo {author}
  {\bibfnamefont {G.}~\bibnamefont {Nobre}}, \bibinfo {author} {\bibfnamefont
  {G.}~\bibnamefont {Noguere}}, \bibinfo {author} {\bibfnamefont
  {M.}~\bibnamefont {Paris}}, \bibinfo {author} {\bibfnamefont
  {M.}~\bibnamefont {Pigni}}, \bibinfo {author} {\bibfnamefont
  {A.}~\bibnamefont {Plompen}}, \bibinfo {author} {\bibfnamefont
  {B.}~\bibnamefont {Pritychenko}}, \bibinfo {author} {\bibfnamefont
  {V.}~\bibnamefont {Pronyaev}}, \bibinfo {author} {\bibfnamefont
  {D.}~\bibnamefont {Roubtsov}}, \bibinfo {author} {\bibfnamefont
  {D.}~\bibnamefont {Rochman}}, \bibinfo {author} {\bibfnamefont
  {P.}~\bibnamefont {Romano}}, \bibinfo {author} {\bibfnamefont
  {P.}~\bibnamefont {Schillebeeckx}}, \bibinfo {author} {\bibfnamefont
  {S.}~\bibnamefont {Simakov}}, \bibinfo {author} {\bibfnamefont
  {M.}~\bibnamefont {Sin}}, \bibinfo {author} {\bibfnamefont {I.}~\bibnamefont
  {Sirakov}}, \bibinfo {author} {\bibfnamefont {B.}~\bibnamefont {Sleaford}},
  \bibinfo {author} {\bibfnamefont {V.}~\bibnamefont {Sobes}}, \bibinfo
  {author} {\bibfnamefont {E.}~\bibnamefont {Soukhovitskii}}, \bibinfo {author}
  {\bibfnamefont {I.}~\bibnamefont {Stetcu}}, \bibinfo {author} {\bibfnamefont
  {P.}~\bibnamefont {Talou}}, \bibinfo {author} {\bibfnamefont
  {I.}~\bibnamefont {Thompson}}, \bibinfo {author} {\bibfnamefont
  {S.}~\bibnamefont {van~der Marck}}, \bibinfo {author} {\bibfnamefont
  {L.}~\bibnamefont {Welser-Sherrill}}, \bibinfo {author} {\bibfnamefont
  {D.}~\bibnamefont {Wiarda}}, \bibinfo {author} {\bibfnamefont
  {M.}~\bibnamefont {White}}, \bibinfo {author} {\bibfnamefont
  {J.}~\bibnamefont {Wormald}}, \bibinfo {author} {\bibfnamefont
  {R.}~\bibnamefont {Wright}}, \bibinfo {author} {\bibfnamefont
  {M.}~\bibnamefont {Zerkle}}, \bibinfo {author} {\bibfnamefont
  {G.}~\bibnamefont {\v{Z}erovnik}},\ and\ \bibinfo {author} {\bibfnamefont
  {Y.}~\bibnamefont {Zhu}},\ }\bibfield  {title} {\bibinfo {title}
  {{ENDF/B-VIII.0}: The {8$^{th}$} major release of the nuclear reaction data
  library with {CIELO}-project cross sections, new standards and thermal
  scattering data},\ }\href
  {https://doi.org/https://doi.org/10.1016/j.nds.2018.02.001} {\bibfield
  {journal} {\bibinfo  {journal} {Nuclear Data Sheets}\ }\textbf {\bibinfo
  {volume} {148}},\ \bibinfo {pages} {1 } (\bibinfo {year} {2018})},\ \bibinfo
  {note} {special Issue on Nuclear Reaction Data}\BibitemShut {NoStop}%
\bibitem [{\citenamefont {Gorley}\ \emph {et~al.}(2020)\citenamefont {Gorley},
  \citenamefont {Diegele}, \citenamefont {Gaganidze}, \citenamefont {Gillemot},
  \citenamefont {Pintsuk}, \citenamefont {Schoofs},\ and\ \citenamefont
  {Szenthe}}]{gorley2020eurofusion}%
  \BibitemOpen
  \bibfield  {author} {\bibinfo {author} {\bibfnamefont {M.}~\bibnamefont
  {Gorley}}, \bibinfo {author} {\bibfnamefont {E.}~\bibnamefont {Diegele}},
  \bibinfo {author} {\bibfnamefont {E.}~\bibnamefont {Gaganidze}}, \bibinfo
  {author} {\bibfnamefont {F.}~\bibnamefont {Gillemot}}, \bibinfo {author}
  {\bibfnamefont {G.}~\bibnamefont {Pintsuk}}, \bibinfo {author} {\bibfnamefont
  {F.}~\bibnamefont {Schoofs}},\ and\ \bibinfo {author} {\bibfnamefont
  {I.}~\bibnamefont {Szenthe}},\ }\bibfield  {title} {\bibinfo {title} {The
  eurofusion materials property handbook for demo in-vessel components---status
  and the challenge to improve confidence level for engineering data},\
  }\href@noop {} {\bibfield  {journal} {\bibinfo  {journal} {Fusion Engineering
  and Design}\ }\textbf {\bibinfo {volume} {158}},\ \bibinfo {pages} {111668}
  (\bibinfo {year} {2020})}\BibitemShut {NoStop}%
\bibitem [{\citenamefont {National Nuclear
  Security~Administration}(2023)}]{NNSA_2023_Mo99_domestic_supply}%
  \BibitemOpen
  \bibfield  {author} {\bibinfo {author} {\bibfnamefont {U.~D. o.~E.}\
  \bibnamefont {National Nuclear Security~Administration}},\ }\href
  {https://www.energy.gov/nnsa/nnsas-molybdenum-99-program-establishing-reliable-domestic-supply-mo-99-without-use-highly}
  {\bibinfo {title} {Nnsa's molybdenum-99 program: Establishing a reliable
  domestic supply of mo-99 without the use of highly enriched uranium}}
  (\bibinfo {year} {2023})\BibitemShut {NoStop}%
\bibitem [{\citenamefont {Schmor}(2011)}]{schmor2011review}%
  \BibitemOpen
  \bibfield  {author} {\bibinfo {author} {\bibfnamefont {P.}~\bibnamefont
  {Schmor}},\ }\bibfield  {title} {\bibinfo {title} {Review of cyclotrons for
  the production of radioactive isotopes for medical and industrial
  applications},\ }\href@noop {} {\bibfield  {journal} {\bibinfo  {journal}
  {Reviews of accelerator science and technology}\ }\textbf {\bibinfo {volume}
  {4}},\ \bibinfo {pages} {103} (\bibinfo {year} {2011})}\BibitemShut {NoStop}%
\bibitem [{\citenamefont {Qaim}(2012)}]{qaim2012present}%
  \BibitemOpen
  \bibfield  {author} {\bibinfo {author} {\bibfnamefont {S.~M.}\ \bibnamefont
  {Qaim}},\ }\bibfield  {title} {\bibinfo {title} {The present and future of
  medical radionuclide production},\ }\href@noop {} {\bibfield  {journal}
  {\bibinfo  {journal} {Radiochimica Acta}\ }\textbf {\bibinfo {volume}
  {100}},\ \bibinfo {pages} {635} (\bibinfo {year} {2012})}\BibitemShut
  {NoStop}%
\bibitem [{\citenamefont {Starovoitova}\ \emph {et~al.}(2014)\citenamefont
  {Starovoitova}, \citenamefont {Tchelidze},\ and\ \citenamefont
  {Wells}}]{starovoitova2014production}%
  \BibitemOpen
  \bibfield  {author} {\bibinfo {author} {\bibfnamefont {V.~N.}\ \bibnamefont
  {Starovoitova}}, \bibinfo {author} {\bibfnamefont {L.}~\bibnamefont
  {Tchelidze}},\ and\ \bibinfo {author} {\bibfnamefont {D.~P.}\ \bibnamefont
  {Wells}},\ }\bibfield  {title} {\bibinfo {title} {Production of medical
  radioisotopes with linear accelerators},\ }\href@noop {} {\bibfield
  {journal} {\bibinfo  {journal} {Applied Radiation and Isotopes}\ }\textbf
  {\bibinfo {volume} {85}},\ \bibinfo {pages} {39} (\bibinfo {year}
  {2014})}\BibitemShut {NoStop}%
\bibitem [{\citenamefont {Wang}\ \emph {et~al.}(2022)\citenamefont {Wang},
  \citenamefont {Chen}, \citenamefont {dos Santos~Augusto}, \citenamefont
  {Liang}, \citenamefont {Qin}, \citenamefont {Liu},\ and\ \citenamefont
  {Liu}}]{wang2022production}%
  \BibitemOpen
  \bibfield  {author} {\bibinfo {author} {\bibfnamefont {Y.}~\bibnamefont
  {Wang}}, \bibinfo {author} {\bibfnamefont {D.}~\bibnamefont {Chen}}, \bibinfo
  {author} {\bibfnamefont {R.}~\bibnamefont {dos Santos~Augusto}}, \bibinfo
  {author} {\bibfnamefont {J.}~\bibnamefont {Liang}}, \bibinfo {author}
  {\bibfnamefont {Z.}~\bibnamefont {Qin}}, \bibinfo {author} {\bibfnamefont
  {J.}~\bibnamefont {Liu}},\ and\ \bibinfo {author} {\bibfnamefont
  {Z.}~\bibnamefont {Liu}},\ }\bibfield  {title} {\bibinfo {title} {Production
  review of accelerator-based medical isotopes},\ }\href@noop {} {\bibfield
  {journal} {\bibinfo  {journal} {Molecules}\ }\textbf {\bibinfo {volume}
  {27}},\ \bibinfo {pages} {5294} (\bibinfo {year} {2022})}\BibitemShut
  {NoStop}%
\bibitem [{\citenamefont {Romano}\ \emph {et~al.}(2015)\citenamefont {Romano},
  \citenamefont {Horelik}, \citenamefont {Herman}, \citenamefont {Nelson},
  \citenamefont {Forget},\ and\ \citenamefont {Smith}}]{romano2015openmc}%
  \BibitemOpen
  \bibfield  {author} {\bibinfo {author} {\bibfnamefont {P.~K.}\ \bibnamefont
  {Romano}}, \bibinfo {author} {\bibfnamefont {N.~E.}\ \bibnamefont {Horelik}},
  \bibinfo {author} {\bibfnamefont {B.~R.}\ \bibnamefont {Herman}}, \bibinfo
  {author} {\bibfnamefont {A.~G.}\ \bibnamefont {Nelson}}, \bibinfo {author}
  {\bibfnamefont {B.}~\bibnamefont {Forget}},\ and\ \bibinfo {author}
  {\bibfnamefont {K.}~\bibnamefont {Smith}},\ }\bibfield  {title} {\bibinfo
  {title} {Openmc: A state-of-the-art monte carlo code for research and
  development},\ }\href@noop {} {\bibfield  {journal} {\bibinfo  {journal}
  {Annals of Nuclear Energy}\ }\textbf {\bibinfo {volume} {82}},\ \bibinfo
  {pages} {90} (\bibinfo {year} {2015})}\BibitemShut {NoStop}%
\bibitem [{\citenamefont {Wong}\ and\ \citenamefont
  {Stambaugh}(2000)}]{wong2000tokamak}%
  \BibitemOpen
  \bibfield  {author} {\bibinfo {author} {\bibfnamefont {C.}~\bibnamefont
  {Wong}}\ and\ \bibinfo {author} {\bibfnamefont {R.}~\bibnamefont
  {Stambaugh}},\ }\bibfield  {title} {\bibinfo {title} {Tokamak reactor designs
  as a function of aspect ratio},\ }\href@noop {} {\bibfield  {journal}
  {\bibinfo  {journal} {Fusion engineering and design}\ }\textbf {\bibinfo
  {volume} {51}},\ \bibinfo {pages} {387} (\bibinfo {year} {2000})}\BibitemShut
  {NoStop}%
\bibitem [{\citenamefont {Bolt}\ \emph {et~al.}(2004)\citenamefont {Bolt},
  \citenamefont {Barabash}, \citenamefont {Krauss}, \citenamefont {Linke},
  \citenamefont {Neu}, \citenamefont {Suzuki}, \citenamefont {Yoshida},
  \citenamefont {Team} \emph {et~al.}}]{bolt2004materials}%
  \BibitemOpen
  \bibfield  {author} {\bibinfo {author} {\bibfnamefont {H.}~\bibnamefont
  {Bolt}}, \bibinfo {author} {\bibfnamefont {V.}~\bibnamefont {Barabash}},
  \bibinfo {author} {\bibfnamefont {W.}~\bibnamefont {Krauss}}, \bibinfo
  {author} {\bibfnamefont {J.}~\bibnamefont {Linke}}, \bibinfo {author}
  {\bibfnamefont {R.}~\bibnamefont {Neu}}, \bibinfo {author} {\bibfnamefont
  {S.}~\bibnamefont {Suzuki}}, \bibinfo {author} {\bibfnamefont
  {N.}~\bibnamefont {Yoshida}}, \bibinfo {author} {\bibfnamefont {A.~U.}\
  \bibnamefont {Team}}, \emph {et~al.},\ }\bibfield  {title} {\bibinfo {title}
  {Materials for the plasma-facing components of fusion reactors},\ }\href@noop
  {} {\bibfield  {journal} {\bibinfo  {journal} {Journal of nuclear materials}\
  }\textbf {\bibinfo {volume} {329}},\ \bibinfo {pages} {66} (\bibinfo {year}
  {2004})}\BibitemShut {NoStop}%
\bibitem [{\citenamefont {Lyon}\ \emph {et~al.}(2008)\citenamefont {Lyon},
  \citenamefont {Ku}, \citenamefont {El-Guebaly}, \citenamefont {Bromberg},
  \citenamefont {Waganer}, \citenamefont {Zarnstorff},\ and\ \citenamefont
  {Team}}]{lyon2008systems}%
  \BibitemOpen
  \bibfield  {author} {\bibinfo {author} {\bibfnamefont {J.}~\bibnamefont
  {Lyon}}, \bibinfo {author} {\bibfnamefont {L.}~\bibnamefont {Ku}}, \bibinfo
  {author} {\bibfnamefont {L.}~\bibnamefont {El-Guebaly}}, \bibinfo {author}
  {\bibfnamefont {L.}~\bibnamefont {Bromberg}}, \bibinfo {author}
  {\bibfnamefont {L.}~\bibnamefont {Waganer}}, \bibinfo {author} {\bibfnamefont
  {M.}~\bibnamefont {Zarnstorff}},\ and\ \bibinfo {author} {\bibfnamefont
  {A.-C.}\ \bibnamefont {Team}},\ }\bibfield  {title} {\bibinfo {title}
  {Systems studies and optimization of the aries-cs power plant},\ }\href@noop
  {} {\bibfield  {journal} {\bibinfo  {journal} {Fusion science and
  technology}\ }\textbf {\bibinfo {volume} {54}},\ \bibinfo {pages} {694}
  (\bibinfo {year} {2008})}\BibitemShut {NoStop}%
\bibitem [{\citenamefont {Meier}\ \emph {et~al.}(2014)\citenamefont {Meier},
  \citenamefont {Dunne}, \citenamefont {Kramer}, \citenamefont {Reyes},
  \citenamefont {Anklam}, \citenamefont {Team} \emph
  {et~al.}}]{meier2014fusion}%
  \BibitemOpen
  \bibfield  {author} {\bibinfo {author} {\bibfnamefont {W.}~\bibnamefont
  {Meier}}, \bibinfo {author} {\bibfnamefont {A.}~\bibnamefont {Dunne}},
  \bibinfo {author} {\bibfnamefont {K.}~\bibnamefont {Kramer}}, \bibinfo
  {author} {\bibfnamefont {S.}~\bibnamefont {Reyes}}, \bibinfo {author}
  {\bibfnamefont {T.}~\bibnamefont {Anklam}}, \bibinfo {author} {\bibfnamefont
  {L.}~\bibnamefont {Team}}, \emph {et~al.},\ }\bibfield  {title} {\bibinfo
  {title} {Fusion technology aspects of laser inertial fusion energy (life)},\
  }\href@noop {} {\bibfield  {journal} {\bibinfo  {journal} {Fusion Engineering
  and Design}\ }\textbf {\bibinfo {volume} {89}},\ \bibinfo {pages} {2489}
  (\bibinfo {year} {2014})}\BibitemShut {NoStop}%
\bibitem [{\citenamefont {Sorbom}\ \emph {et~al.}(2015)\citenamefont {Sorbom},
  \citenamefont {Ball}, \citenamefont {Palmer}, \citenamefont {Mangiarotti},
  \citenamefont {Sierchio}, \citenamefont {Bonoli}, \citenamefont {Kasten},
  \citenamefont {Sutherland}, \citenamefont {Barnard}, \citenamefont
  {Haakonsen}, \citenamefont {Goh}, \citenamefont {Sung},\ and\ \citenamefont
  {Whyte}}]{Sorbom2015}%
  \BibitemOpen
  \bibfield  {author} {\bibinfo {author} {\bibfnamefont {B.~N.}\ \bibnamefont
  {Sorbom}}, \bibinfo {author} {\bibfnamefont {J.}~\bibnamefont {Ball}},
  \bibinfo {author} {\bibfnamefont {T.~R.}\ \bibnamefont {Palmer}}, \bibinfo
  {author} {\bibfnamefont {F.~J.}\ \bibnamefont {Mangiarotti}}, \bibinfo
  {author} {\bibfnamefont {J.~M.}\ \bibnamefont {Sierchio}}, \bibinfo {author}
  {\bibfnamefont {P.}~\bibnamefont {Bonoli}}, \bibinfo {author} {\bibfnamefont
  {C.}~\bibnamefont {Kasten}}, \bibinfo {author} {\bibfnamefont {D.~A.}\
  \bibnamefont {Sutherland}}, \bibinfo {author} {\bibfnamefont {H.~S.}\
  \bibnamefont {Barnard}}, \bibinfo {author} {\bibfnamefont {C.~B.}\
  \bibnamefont {Haakonsen}}, \bibinfo {author} {\bibfnamefont {J.}~\bibnamefont
  {Goh}}, \bibinfo {author} {\bibfnamefont {C.}~\bibnamefont {Sung}},\ and\
  \bibinfo {author} {\bibfnamefont {D.~G.}\ \bibnamefont {Whyte}},\ }\bibfield
  {title} {\bibinfo {title} {{ARC: A compact, high-field, fusion nuclear
  science facility and demonstration power plant with demountable magnets}},\
  }\href@noop {} {\bibfield  {journal} {\bibinfo  {journal} {Fusion Engineering
  and Design}\ }\textbf {\bibinfo {volume} {100}},\ \bibinfo {pages} {378}
  (\bibinfo {year} {2015})}\BibitemShut {NoStop}%
\bibitem [{\citenamefont {Park}(2000)}]{IAEA_INIS_P32_2000}%
  \BibitemOpen
  \bibfield  {author} {\bibinfo {author} {\bibfnamefont {U.}~\bibnamefont
  {Park}},\ }\href
  {https://inis.iaea.org/collection/NCLCollectionStore/_Public/32/065/32065604.pdf}
  {\emph {\bibinfo {title} {Carrier-free $^{32}$P production via
  $^{32}$S(n,p)$^{32}$P}}},\ \bibinfo {type} {Tech. Rep.}\ (\bibinfo
  {institution} {IAEA INIS},\ \bibinfo {year} {2000})\BibitemShut {NoStop}%
\bibitem [{INI(2013)}]{INIS_P32_Assessment_2013}%
  \BibitemOpen
  \href
  {https://www.researchgate.net/publication/261363150_Reactor_production_of_32P_for_medical_applications_An_assessment_of_32Snp32P_and_31Png_32P_methods}
  {\bibinfo {title} {Reactor production of $^{32}$p: assessment of
  $^{32}$s(n,p) and $^{31}$p(n,$\gamma$)}},\ \bibinfo {howpublished}
  {ResearchGate preprint} (\bibinfo {year} {2013})\BibitemShut {NoStop}%
\bibitem [{\citenamefont {Azarov}(2021)}]{Azarov2021_P32_NG}%
  \BibitemOpen
  \bibfield  {author} {\bibinfo {author} {\bibfnamefont {O.}~\bibnamefont
  {Azarov}},\ }\href {https://inis.iaea.org/records/2ysw9-cmj04} {\bibinfo
  {title} {Production of phosphorus-32 at a neutron generator}},\ \bibinfo
  {howpublished} {IAEA INIS} (\bibinfo {year} {2021})\BibitemShut {NoStop}%
\bibitem [{\citenamefont {Pantel}\ \emph {et~al.}(2023)\citenamefont {Pantel},
  \citenamefont {Eiber}, \citenamefont {Beyder}, \citenamefont {Kendi},
  \citenamefont {Laforest}, \citenamefont {Rauscher}, \citenamefont
  {Silberstein},\ and\ \citenamefont {Thorpe}}]{Pantel2023}%
  \BibitemOpen
  \bibfield  {author} {\bibinfo {author} {\bibfnamefont {A.~R.}\ \bibnamefont
  {Pantel}}, \bibinfo {author} {\bibfnamefont {M.}~\bibnamefont {Eiber}},
  \bibinfo {author} {\bibfnamefont {D.~D.}\ \bibnamefont {Beyder}}, \bibinfo
  {author} {\bibfnamefont {A.~T.}\ \bibnamefont {Kendi}}, \bibinfo {author}
  {\bibfnamefont {R.}~\bibnamefont {Laforest}}, \bibinfo {author}
  {\bibfnamefont {I.}~\bibnamefont {Rauscher}}, \bibinfo {author}
  {\bibfnamefont {E.~B.}\ \bibnamefont {Silberstein}},\ and\ \bibinfo {author}
  {\bibfnamefont {M.~P.}\ \bibnamefont {Thorpe}},\ }\bibfield  {title}
  {\bibinfo {title} {{SNMMI} procedure standard/{EANM} practice guideline for
  palliative nuclear medicine therapies of bone metastases},\ }\href
  {https://doi.org/10.2967/jnmt.123.265936} {\bibfield  {journal} {\bibinfo
  {journal} {J.~Nucl.~Med.~Technol.}\ }\textbf {\bibinfo {volume} {51}},\
  \bibinfo {pages} {176} (\bibinfo {year} {2023})}\BibitemShut {NoStop}%
\bibitem [{\citenamefont {Dash}\ and\ \citenamefont {Das}(2025)}]{Dash2025}%
  \BibitemOpen
  \bibfield  {author} {\bibinfo {author} {\bibfnamefont {A.}~\bibnamefont
  {Dash}}\ and\ \bibinfo {author} {\bibfnamefont {T.}~\bibnamefont {Das}},\
  }\bibfield  {title} {\bibinfo {title} {Production of therapeutic
  radionuclides for bone-seeking radiopharmaceuticals to meet clinical exigency
  and academic research},\ }\href {https://doi.org/10.1007/s10967-025-10025-1}
  {\bibfield  {journal} {\bibinfo  {journal} {J.~Radioanal.~Nucl.~Chem.}\
  }\textbf {\bibinfo {volume} {334}},\ \bibinfo {pages} {2591} (\bibinfo {year}
  {2025})}\BibitemShut {NoStop}%
\bibitem [{\citenamefont {Van~de Voorde}\ and\ \citenamefont
  {et~al.}(2021)}]{Voorde2021_Frontiers_Sm153HSA}%
  \BibitemOpen
  \bibfield  {author} {\bibinfo {author} {\bibfnamefont {M.}~\bibnamefont
  {Van~de Voorde}}\ and\ \bibinfo {author} {\bibnamefont {et~al.}},\ }\bibfield
   {title} {\bibinfo {title} {Production of $^{153}$sm with very high specific
  activity for trnt},\ }\href
  {https://www.frontiersin.org/articles/10.3389/fmed.2021.675221/full}
  {\bibfield  {journal} {\bibinfo  {journal} {Front. Med.}\ } (\bibinfo {year}
  {2021})}\BibitemShut {NoStop}%
\bibitem [{\citenamefont {Vermeulen}\ and\ \citenamefont
  {et~al.}(2022)}]{Vermeulen2022_Sm153_MassSep}%
  \BibitemOpen
  \bibfield  {author} {\bibinfo {author} {\bibfnamefont {K.}~\bibnamefont
  {Vermeulen}}\ and\ \bibinfo {author} {\bibnamefont {et~al.}},\ }\bibfield
  {title} {\bibinfo {title} {High-molar-activity $^{153}$sm via mass
  separation},\ }\href {https://pmc.ncbi.nlm.nih.gov/articles/PMC9785812/}
  {\bibfield  {journal} {\bibinfo  {journal} {Eur. J. Nucl. Med. Mol. Imaging}\
  } (\bibinfo {year} {2022})}\BibitemShut {NoStop}%
\bibitem [{\citenamefont {Weber}\ \emph {et~al.}(2022)\citenamefont {Weber},
  \citenamefont {Lam}, \citenamefont {Chiesa}, \citenamefont {Konijnenberg},
  \citenamefont {Cremonesi}, \citenamefont {Flamen}, \citenamefont {Gnesin},
  \citenamefont {Bodei}, \citenamefont {Kracmerova}, \citenamefont {Luster},
  \citenamefont {Garin},\ and\ \citenamefont {Herrmann}}]{Weber2022}%
  \BibitemOpen
  \bibfield  {author} {\bibinfo {author} {\bibfnamefont {M.}~\bibnamefont
  {Weber}}, \bibinfo {author} {\bibfnamefont {M.}~\bibnamefont {Lam}}, \bibinfo
  {author} {\bibfnamefont {C.}~\bibnamefont {Chiesa}}, \bibinfo {author}
  {\bibfnamefont {M.}~\bibnamefont {Konijnenberg}}, \bibinfo {author}
  {\bibfnamefont {M.}~\bibnamefont {Cremonesi}}, \bibinfo {author}
  {\bibfnamefont {P.}~\bibnamefont {Flamen}}, \bibinfo {author} {\bibfnamefont
  {S.}~\bibnamefont {Gnesin}}, \bibinfo {author} {\bibfnamefont
  {L.}~\bibnamefont {Bodei}}, \bibinfo {author} {\bibfnamefont
  {T.}~\bibnamefont {Kracmerova}}, \bibinfo {author} {\bibfnamefont
  {M.}~\bibnamefont {Luster}}, \bibinfo {author} {\bibfnamefont
  {E.}~\bibnamefont {Garin}},\ and\ \bibinfo {author} {\bibfnamefont
  {K.}~\bibnamefont {Herrmann}},\ }\bibfield  {title} {\bibinfo {title} {{EANM}
  procedure guideline for the treatment of liver cancer and liver metastases
  with intra-arterial radioactive compounds},\ }\href
  {https://doi.org/10.1007/s00259-021-05600-z} {\bibfield  {journal} {\bibinfo
  {journal} {Eur.~J.~Nucl.~Med.~Mol.~Imaging}\ }\textbf {\bibinfo {volume}
  {49}},\ \bibinfo {pages} {1682} (\bibinfo {year} {2022})}\BibitemShut
  {NoStop}%
\bibitem [{\citenamefont {Tennvall}\ \emph {et~al.}(2007)\citenamefont
  {Tennvall}, \citenamefont {Fischer}, \citenamefont {Bischof~Delaloye},
  \citenamefont {Bombardieri}, \citenamefont {Bodei}, \citenamefont
  {Giammarile}, \citenamefont {Lassmann}, \citenamefont {Oyen},\ and\
  \citenamefont {Brans}}]{Tennvall2007}%
  \BibitemOpen
  \bibfield  {author} {\bibinfo {author} {\bibfnamefont {J.}~\bibnamefont
  {Tennvall}}, \bibinfo {author} {\bibfnamefont {M.}~\bibnamefont {Fischer}},
  \bibinfo {author} {\bibfnamefont {A.}~\bibnamefont {Bischof~Delaloye}},
  \bibinfo {author} {\bibfnamefont {E.}~\bibnamefont {Bombardieri}}, \bibinfo
  {author} {\bibfnamefont {L.}~\bibnamefont {Bodei}}, \bibinfo {author}
  {\bibfnamefont {F.}~\bibnamefont {Giammarile}}, \bibinfo {author}
  {\bibfnamefont {M.}~\bibnamefont {Lassmann}}, \bibinfo {author}
  {\bibfnamefont {W.}~\bibnamefont {Oyen}},\ and\ \bibinfo {author}
  {\bibfnamefont {B.}~\bibnamefont {Brans}},\ }\bibfield  {title} {\bibinfo
  {title} {{EANM} procedure guideline for radio-immunotherapy of b-cell
  lymphoma with $^{90}$y-radiolabelled ibritumomab tiuxetan
  (zevalin{\textregistered})},\ }\href
  {https://doi.org/10.1007/s00259-007-0365-8} {\bibfield  {journal} {\bibinfo
  {journal} {Eur.~J.~Nucl.~Med.~Mol.~Imaging}\ }\textbf {\bibinfo {volume}
  {34}},\ \bibinfo {pages} {616} (\bibinfo {year} {2007})}\BibitemShut
  {NoStop}%
\bibitem [{\citenamefont {Basu}\ and\ \citenamefont
  {McCutchan}(2020)}]{Basu2020}%
  \BibitemOpen
  \bibfield  {author} {\bibinfo {author} {\bibfnamefont {S.~K.}\ \bibnamefont
  {Basu}}\ and\ \bibinfo {author} {\bibfnamefont {E.~A.}\ \bibnamefont
  {McCutchan}},\ }\bibfield  {title} {\bibinfo {title} {Nuclear data sheets for
  a = 90},\ }\href {https://doi.org/10.1016/j.nds.2020.03.001} {\bibfield
  {journal} {\bibinfo  {journal} {Nucl.~Data~Sheets}\ }\textbf {\bibinfo
  {volume} {165}},\ \bibinfo {pages} {1} (\bibinfo {year} {2020})}\BibitemShut
  {NoStop}%
\bibitem [{\citenamefont {Vogel}\ and\ \citenamefont
  {et~al.}(2021)}]{Vogel2021_Lu177_Review}%
  \BibitemOpen
  \bibfield  {author} {\bibinfo {author} {\bibfnamefont {W.}~\bibnamefont
  {Vogel}}\ and\ \bibinfo {author} {\bibnamefont {et~al.}},\ }\bibfield
  {title} {\bibinfo {title} {Challenges and future options for the production
  of lutetium-177},\ }\href {https://pmc.ncbi.nlm.nih.gov/articles/PMC8241800/}
  {\bibfield  {journal} {\bibinfo  {journal} {EJNMMI Radiopharm. Chem.}\ }
  (\bibinfo {year} {2021})}\BibitemShut {NoStop}%
\bibitem [{\citenamefont {Morris}\ and\ \citenamefont
  {et~al.}(2025)}]{Morris2025_CurrMedChem_DeuteronLu177}%
  \BibitemOpen
  \bibfield  {author} {\bibinfo {author} {\bibfnamefont {A.}~\bibnamefont
  {Morris}}\ and\ \bibinfo {author} {\bibnamefont {et~al.}},\ }\bibfield
  {title} {\bibinfo {title} {Compact accelerator-based production of
  carrier-free $^{177}$lu},\ }\href
  {https://pmc.ncbi.nlm.nih.gov/articles/PMC12379675/} {\bibfield  {journal}
  {\bibinfo  {journal} {Curr. Med. Chem.}\ } (\bibinfo {year}
  {2025})}\BibitemShut {NoStop}%
\bibitem [{\citenamefont {Vosoughi}\ \emph {et~al.}(2017)\citenamefont
  {Vosoughi}, \citenamefont {Saidi}, \citenamefont {Jafarizadeh}, \citenamefont
  {Aslani},\ and\ \citenamefont {Ansari}}]{Vosoughi2017_IJNMM_Dy_Ho}%
  \BibitemOpen
  \bibfield  {author} {\bibinfo {author} {\bibfnamefont {A.}~\bibnamefont
  {Vosoughi}}, \bibinfo {author} {\bibfnamefont {K.}~\bibnamefont {Saidi}},
  \bibinfo {author} {\bibfnamefont {M.}~\bibnamefont {Jafarizadeh}}, \bibinfo
  {author} {\bibfnamefont {G.}~\bibnamefont {Aslani}},\ and\ \bibinfo {author}
  {\bibfnamefont {S.}~\bibnamefont {Ansari}},\ }\bibfield  {title} {\bibinfo
  {title} {Production routes and specific activity of ${}^{166}$ho for
  therapeutic applications},\ }\href
  {https://doi.org/10.1007/s13226-017-0416-7} {\bibfield  {journal} {\bibinfo
  {journal} {Int. J. Nucl. Med. Mol. Imaging}\ }\textbf {\bibinfo {volume}
  {30}},\ \bibinfo {pages} {43} (\bibinfo {year} {2017})}\BibitemShut {NoStop}%
\bibitem [{\citenamefont {Zimmerman}\ \emph {et~al.}(2024)\citenamefont
  {Zimmerman}, \citenamefont {Woods},\ and\ \citenamefont {van~der
  Meulen}}]{Zimmerman2024_DDEP_166Ho}%
  \BibitemOpen
  \bibfield  {author} {\bibinfo {author} {\bibfnamefont {B.~E.}\ \bibnamefont
  {Zimmerman}}, \bibinfo {author} {\bibfnamefont {M.~J.}\ \bibnamefont
  {Woods}},\ and\ \bibinfo {author} {\bibfnamefont {N.~P.}\ \bibnamefont
  {van~der Meulen}},\ }\bibfield  {title} {\bibinfo {title} {Recommended data
  for the decay of ${}^{166}$ho for nuclear medicine applications},\ }\href
  {https://doi.org/10.1016/j.apradiso.2024.111234} {\bibfield  {journal}
  {\bibinfo  {journal} {Appl. Radiat. Isot.}\ }\textbf {\bibinfo {volume}
  {212}},\ \bibinfo {pages} {111234} (\bibinfo {year} {2024})}\BibitemShut
  {NoStop}%
\bibitem [{\citenamefont {Prince}\ \emph {et~al.}(2018)\citenamefont {Prince},
  \citenamefont {Bruijnen}, \citenamefont {van Rooij}, \citenamefont {Braat},
  \citenamefont {Smits}, \citenamefont {Bastiaannet}, \citenamefont {de~Jong},\
  and\ \citenamefont {Lam}}]{Prince2018_JNM_Ho166}%
  \BibitemOpen
  \bibfield  {author} {\bibinfo {author} {\bibfnamefont {J.~F.}\ \bibnamefont
  {Prince}}, \bibinfo {author} {\bibfnamefont {R.~C.~G.}\ \bibnamefont
  {Bruijnen}}, \bibinfo {author} {\bibfnamefont {R.}~\bibnamefont {van Rooij}},
  \bibinfo {author} {\bibfnamefont {A.~J. A.~T.}\ \bibnamefont {Braat}},
  \bibinfo {author} {\bibfnamefont {M.~L.~J.}\ \bibnamefont {Smits}}, \bibinfo
  {author} {\bibfnamefont {R.}~\bibnamefont {Bastiaannet}}, \bibinfo {author}
  {\bibfnamefont {H.~W. A.~M.}\ \bibnamefont {de~Jong}},\ and\ \bibinfo
  {author} {\bibfnamefont {M.~G. E.~H.}\ \bibnamefont {Lam}},\ }\bibfield
  {title} {\bibinfo {title} {Safety of ${}^{166}$ho radioembolization in
  patients with advanced liver tumors: The hepar~1 study},\ }\href
  {https://doi.org/10.2967/jnumed.117.198143} {\bibfield  {journal} {\bibinfo
  {journal} {J. Nucl. Med.}\ }\textbf {\bibinfo {volume} {59}},\ \bibinfo
  {pages} {659} (\bibinfo {year} {2018})}\BibitemShut {NoStop}%
\bibitem [{\citenamefont {Klaassen}\ \emph {et~al.}(2019)\citenamefont
  {Klaassen}, \citenamefont {van Es}, \citenamefont {de~Wit-van~der Veen},\
  and\ \citenamefont {Nijsen}}]{Klaassen2019_EJNMMIPharmChem_Ho166}%
  \BibitemOpen
  \bibfield  {author} {\bibinfo {author} {\bibfnamefont {N.~J.~M.}\
  \bibnamefont {Klaassen}}, \bibinfo {author} {\bibfnamefont {R.~J.~J.}\
  \bibnamefont {van Es}}, \bibinfo {author} {\bibfnamefont {L.~J.}\
  \bibnamefont {de~Wit-van~der Veen}},\ and\ \bibinfo {author} {\bibfnamefont
  {J.~F.~W.}\ \bibnamefont {Nijsen}},\ }\bibfield  {title} {\bibinfo {title}
  {Holmium-166 therapeutic radionuclide: properties and clinical potential},\
  }\href {https://doi.org/10.1186/s41181-019-0077-0} {\bibfield  {journal}
  {\bibinfo  {journal} {EJNMMI Radiopharm. Chem.}\ }\textbf {\bibinfo {volume}
  {4}},\ \bibinfo {pages} {19} (\bibinfo {year} {2019})}\BibitemShut {NoStop}%
\bibitem [{\citenamefont {Ntihabose}\ and\ \citenamefont
  {et~al.}(2025)}]{Ntihabose2025_EJNMMIPharmChem_Tb161Review}%
  \BibitemOpen
  \bibfield  {author} {\bibinfo {author} {\bibfnamefont {C.}~\bibnamefont
  {Ntihabose}}\ and\ \bibinfo {author} {\bibnamefont {et~al.}},\ }\bibfield
  {title} {\bibinfo {title} {Potentials and practical challenges of terbium-161
  labeled radiopharmaceuticals},\ }\href
  {https://pmc.ncbi.nlm.nih.gov/articles/PMC12494511/} {\bibfield  {journal}
  {\bibinfo  {journal} {EJNMMI Radiopharm. Chem.}\ } (\bibinfo {year}
  {2025})}\BibitemShut {NoStop}%
\bibitem [{\citenamefont {Santos}\ and\ \citenamefont
  {et~al.}(2025)}]{Santos2025_Tb161_Mito}%
  \BibitemOpen
  \bibfield  {author} {\bibinfo {author} {\bibfnamefont {J.}~\bibnamefont
  {Santos}}\ and\ \bibinfo {author} {\bibnamefont {et~al.}},\ }\bibfield
  {title} {\bibinfo {title} {Mitochondria-tropic $^{161}$tb radiocomplexes},\
  }\href
  {https://ejnmmipharmchem.springeropen.com/articles/10.1186/s41181-025-00339-6}
  {\bibfield  {journal} {\bibinfo  {journal} {EJNMMI Radiopharm. Chem.}\ }
  (\bibinfo {year} {2025})}\BibitemShut {NoStop}%
\bibitem [{\citenamefont {Kondev}\ \emph {et~al.}(2018)\citenamefont {Kondev},
  \citenamefont {Juutinen},\ and\ \citenamefont {Hartley}}]{Kondev2018}%
  \BibitemOpen
  \bibfield  {author} {\bibinfo {author} {\bibfnamefont {F.~G.}\ \bibnamefont
  {Kondev}}, \bibinfo {author} {\bibfnamefont {S.}~\bibnamefont {Juutinen}},\
  and\ \bibinfo {author} {\bibfnamefont {D.~J.}\ \bibnamefont {Hartley}},\
  }\bibfield  {title} {\bibinfo {title} {Nuclear data sheets for a = 188},\
  }\href {https://doi.org/10.1016/j.nds.2018.05.001} {\bibfield  {journal}
  {\bibinfo  {journal} {Nucl.~Data~Sheets}\ }\textbf {\bibinfo {volume}
  {150}},\ \bibinfo {pages} {1} (\bibinfo {year} {2018})}\BibitemShut {NoStop}%
\bibitem [{\citenamefont {Zhou}\ and\ \citenamefont
  {et~al.}(2019)}]{Zhou2019_Cu64_Review}%
  \BibitemOpen
  \bibfield  {author} {\bibinfo {author} {\bibfnamefont {Y.}~\bibnamefont
  {Zhou}}\ and\ \bibinfo {author} {\bibnamefont {et~al.}},\ }\bibfield  {title}
  {\bibinfo {title} {$^{64}$cu radiopharmaceuticals in molecular imaging},\
  }\href {https://pmc.ncbi.nlm.nih.gov/articles/PMC6378420/} {\bibfield
  {journal} {\bibinfo  {journal} {Curr. Top. Med. Chem.}\ } (\bibinfo {year}
  {2019})}\BibitemShut {NoStop}%
\bibitem [{\citenamefont {Svedjehed}\ and\ \citenamefont
  {et~al.}(2020)}]{Svedjehed2020_EJNMMIPharmChem}%
  \BibitemOpen
  \bibfield  {author} {\bibinfo {author} {\bibfnamefont {J.}~\bibnamefont
  {Svedjehed}}\ and\ \bibinfo {author} {\bibnamefont {et~al.}},\ }\bibfield
  {title} {\bibinfo {title} {Automated isolation of high-purity $^{61/64}$cu
  (applicable to $^{64}$ni(p,n)$^{64}$cu)},\ }\href
  {https://ejnmmipharmchem.springeropen.com/articles/10.1186/s41181-020-00108-7}
  {\bibfield  {journal} {\bibinfo  {journal} {EJNMMI Radiopharm. Chem.}\ }
  (\bibinfo {year} {2020})}\BibitemShut {NoStop}%
\bibitem [{IAE(2019)}]{IAEA_2019_Cu64_Brochure}%
  \BibitemOpen
  \href {https://www-pub.iaea.org/MTCD/Publications/PDF/PUB1961_web.pdf}
  {\bibinfo {title} {Copper-64 radiopharmaceuticals: production and qc}},\
  \bibinfo {howpublished} {IAEA} (\bibinfo {year} {2019})\BibitemShut {NoStop}%
\bibitem [{\citenamefont {Krasnovskaya}\ and\ \citenamefont
  {et~al.}(2023)}]{Krasnovskaya2023_Cu64_67_Review}%
  \BibitemOpen
  \bibfield  {author} {\bibinfo {author} {\bibfnamefont {O.}~\bibnamefont
  {Krasnovskaya}}\ and\ \bibinfo {author} {\bibnamefont {et~al.}},\ }\bibfield
  {title} {\bibinfo {title} {Recent advances in $^{64}$cu/$^{67}$cu-based
  theranostics},\ }\href {https://www.mdpi.com/1422-0067/24/11/9154} {\bibfield
   {journal} {\bibinfo  {journal} {Int. J. Mol. Sci.}\ } (\bibinfo {year}
  {2023})}\BibitemShut {NoStop}%
\bibitem [{\citenamefont {Mohammed}\ and\ \citenamefont
  {et~al.}(2020)}]{Mohammed2020_AIP_IndiumRoutes}%
  \BibitemOpen
  \bibfield  {author} {\bibinfo {author} {\bibfnamefont {R.}~\bibnamefont
  {Mohammed}}\ and\ \bibinfo {author} {\bibnamefont {et~al.}},\ }\bibfield
  {title} {\bibinfo {title} {Evaluation of cross-sections for indium medical
  isotopes},\ }in\ \href {https://doi.org/10.1063/5.0028096} {\emph {\bibinfo
  {booktitle} {AIP Conf. Proc.}}}\ (\bibinfo {year} {2020})\BibitemShut
  {NoStop}%
\bibitem [{Ind(2025)}]{Indium111_TopicPage_Elsevier}%
  \BibitemOpen
  \href
  {https://www.sciencedirect.com/topics/medicine-and-dentistry/indium-111}
  {\bibinfo {title} {Indium-111: production and applications (topic
  overview)}},\ \bibinfo {howpublished} {Elsevier Topic Pages} (\bibinfo {year}
  {2025})\BibitemShut {NoStop}%
\bibitem [{\citenamefont {Avram}\ and\ \citenamefont
  {et~al.}(2022)}]{Avram2022_JNM_Guideline}%
  \BibitemOpen
  \bibfield  {author} {\bibinfo {author} {\bibfnamefont {A.}~\bibnamefont
  {Avram}}\ and\ \bibinfo {author} {\bibnamefont {et~al.}},\ }\bibfield
  {title} {\bibinfo {title} {Snmmi procedure standard/eanm practice guideline
  for treatment of differentiated thyroid cancer with $^{131}$i},\ }\href
  {https://pure.eur.nl/files/137006590/15N.full.pdf} {\bibfield  {journal}
  {\bibinfo  {journal} {J. Nucl. Med.}\ } (\bibinfo {year} {2022})}\BibitemShut
  {NoStop}%
\bibitem [{\citenamefont {Petranovi{\'c}~Ov{\v{c}}ari{\v{c}}ek}\ and\
  \citenamefont {et~al.}(2022)}]{Petranovic2022_EJNMMI_ETA}%
  \BibitemOpen
  \bibfield  {author} {\bibinfo {author} {\bibfnamefont {P.}~\bibnamefont
  {Petranovi{\'c}~Ov{\v{c}}ari{\v{c}}ek}}\ and\ \bibinfo {author} {\bibnamefont
  {et~al.}},\ }\bibfield  {title} {\bibinfo {title} {Snmmi/eanm guideline vs.
  eta statement in dtc management},\ }\href
  {https://doi.org/10.1007/s00259-022-05935-1} {\bibfield  {journal} {\bibinfo
  {journal} {Eur. J. Nucl. Med. Mol. Imaging}\ }\textbf {\bibinfo {volume}
  {49}},\ \bibinfo {pages} {3959} (\bibinfo {year} {2022})}\BibitemShut
  {NoStop}%
\bibitem [{\citenamefont {Mishra}\ and\ \citenamefont
  {et~al.}(2021)}]{Mishra2021_ARI_131IYield}%
  \BibitemOpen
  \bibfield  {author} {\bibinfo {author} {\bibfnamefont {A.}~\bibnamefont
  {Mishra}}\ and\ \bibinfo {author} {\bibnamefont {et~al.}},\ }\bibfield
  {title} {\bibinfo {title} {Estimation and verification of $^{131}$i yield
  from fission and $^{130}$te(n,$\gamma$)},\ }\bibfield  {journal} {\bibinfo
  {journal} {Appl. Radiat. Isot.}\ }\href
  {https://doi.org/10.1016/j.apradiso.2020.109290}
  {10.1016/j.apradiso.2020.109290} (\bibinfo {year} {2021})\BibitemShut
  {NoStop}%
\bibitem [{\citenamefont {Chattopadhyay}\ and\ \citenamefont
  {et~al.}(2009)}]{Chattopadhyay2009_ARI_131I_Separation}%
  \BibitemOpen
  \bibfield  {author} {\bibinfo {author} {\bibfnamefont {S.}~\bibnamefont
  {Chattopadhyay}}\ and\ \bibinfo {author} {\bibnamefont {et~al.}},\ }\bibfield
   {title} {\bibinfo {title} {Rapid separation of no-carrier-added $^{131}$i
  from irradiated teo$_2$},\ }\bibfield  {journal} {\bibinfo  {journal} {Appl.
  Radiat. Isot.}\ }\href {https://doi.org/10.1016/j.apradiso.2009.03.006}
  {10.1016/j.apradiso.2009.03.006} (\bibinfo {year} {2009})\BibitemShut
  {NoStop}%
\bibitem [{\citenamefont {Hilgers}\ \emph {et~al.}(2008)\citenamefont
  {Hilgers}, \citenamefont {Coenen},\ and\ \citenamefont
  {Qaim}}]{Hilgers2008Pt195mAlpha}%
  \BibitemOpen
  \bibfield  {author} {\bibinfo {author} {\bibfnamefont {K.}~\bibnamefont
  {Hilgers}}, \bibinfo {author} {\bibfnamefont {H.}~\bibnamefont {Coenen}},\
  and\ \bibinfo {author} {\bibfnamefont {S.}~\bibnamefont {Qaim}},\ }\bibfield
  {title} {\bibinfo {title} {Production of $^{193m}$pt and $^{195m}$pt with
  high specific activity via $\alpha$ on $^{192}$os},\ }\bibfield  {journal}
  {\bibinfo  {journal} {Appl. Radiat. Isot.}\ }\href
  {https://doi.org/10.1016/j.apradiso.2007.10.001}
  {10.1016/j.apradiso.2007.10.001} (\bibinfo {year} {2008})\BibitemShut
  {NoStop}%
\bibitem [{\citenamefont {Knapp}\ \emph {et~al.}(2005)\citenamefont {Knapp},
  \citenamefont {Mirzadeh}, \citenamefont {Beets},\ and\ \citenamefont
  {Du}}]{Knapp2005_HFIR_Isotopes}%
  \BibitemOpen
  \bibfield  {author} {\bibinfo {author} {\bibfnamefont {F.}~\bibnamefont
  {Knapp}}, \bibinfo {author} {\bibfnamefont {S.}~\bibnamefont {Mirzadeh}},
  \bibinfo {author} {\bibfnamefont {A.}~\bibnamefont {Beets}},\ and\ \bibinfo
  {author} {\bibfnamefont {M.}~\bibnamefont {Du}},\ }\bibfield  {title}
  {\bibinfo {title} {Production of therapeutic radioisotopes in ornl hfir},\
  }\href {https://doi.org/10.1007/s10967-005-0083-4} {\bibfield  {journal}
  {\bibinfo  {journal} {J. Radioanal. Nucl. Chem.}\ }\textbf {\bibinfo {volume}
  {263}},\ \bibinfo {pages} {503} (\bibinfo {year} {2005})}\BibitemShut
  {NoStop}%
\bibitem [{\citenamefont {Madumarov}\ and\ \citenamefont
  {et~al.}(2024)}]{Madumarov2024_Pt195m_DoubleCap}%
  \BibitemOpen
  \bibfield  {author} {\bibinfo {author} {\bibfnamefont {A.}~\bibnamefont
  {Madumarov}}\ and\ \bibinfo {author} {\bibnamefont {et~al.}},\ }\bibfield
  {title} {\bibinfo {title} {Double neutron capture on $^{193}$ir for
  $^{195m}$pt via $^{195m}$ir},\ }\bibfield  {journal} {\bibinfo  {journal}
  {Appl. Radiat. Isot.}\ }\href
  {https://doi.org/10.1016/j.apradiso.2024.110409}
  {10.1016/j.apradiso.2024.110409} (\bibinfo {year} {2024})\BibitemShut
  {NoStop}%
\bibitem [{\citenamefont {de~Roest}\ and\ \citenamefont
  {et~al.}(2024)}]{deRoest2024_Pt195m_PD}%
  \BibitemOpen
  \bibfield  {author} {\bibinfo {author} {\bibfnamefont {R.}~\bibnamefont
  {de~Roest}}\ and\ \bibinfo {author} {\bibnamefont {et~al.}},\ }\bibfield
  {title} {\bibinfo {title} {Pharmacodynamics and biodistribution of
  [${^{195m}}$pt]cisplatin},\ }\href
  {https://ejnmmires.springeropen.com/articles/10.1186/s13550-024-01082-w}
  {\bibfield  {journal} {\bibinfo  {journal} {EJNMMI Res.}\ } (\bibinfo {year}
  {2024})}\BibitemShut {NoStop}%
\bibitem [{\citenamefont {Hoogenkamp}\ and\ \citenamefont
  {et~al.}(2025)}]{Hoogenkamp2025_Pt195m_Clinical}%
  \BibitemOpen
  \bibfield  {author} {\bibinfo {author} {\bibfnamefont {D.}~\bibnamefont
  {Hoogenkamp}}\ and\ \bibinfo {author} {\bibnamefont {et~al.}},\ }\bibfield
  {title} {\bibinfo {title} {[$^{195m}$pt]cisplatin for lung cancer imaging: a
  pilot study},\ }\href
  {https://ejnmmires.springeropen.com/articles/10.1186/s13550-025-01281-z}
  {\bibfield  {journal} {\bibinfo  {journal} {EJNMMI Res.}\ } (\bibinfo {year}
  {2025})}\BibitemShut {NoStop}%
\bibitem [{\citenamefont {Saidi}\ and\ \citenamefont
  {et~al.}(2011)}]{Saidi2011_Pd103_Rhpn}%
  \BibitemOpen
  \bibfield  {author} {\bibinfo {author} {\bibfnamefont {P.}~\bibnamefont
  {Saidi}}\ and\ \bibinfo {author} {\bibnamefont {et~al.}},\ }\bibfield
  {title} {\bibinfo {title} {Cyclotron production of $^{103}$pd via
  $^{103}$rh(p,n)},\ }\bibfield  {journal} {\bibinfo  {journal} {Nucl. Instrum.
  Methods Phys. Res. B}\ }\href {https://doi.org/10.1016/j.nimb.2011.02.019}
  {10.1016/j.nimb.2011.02.019} (\bibinfo {year} {2011})\BibitemShut {NoStop}%
\bibitem [{\citenamefont {{\"U}nc{\"u}}\ and\ \citenamefont
  {et~al.}(2023)}]{Uncu2023_Pd103_Routes}%
  \BibitemOpen
  \bibfield  {author} {\bibinfo {author} {\bibfnamefont {Y.}~\bibnamefont
  {{\"U}nc{\"u}}}\ and\ \bibinfo {author} {\bibnamefont {et~al.}},\ }\bibfield
  {title} {\bibinfo {title} {Production routes of $^{103}$pd in
  no-carrier-added form},\ }\bibfield  {journal} {\bibinfo  {journal} {Appl.
  Radiat. Isot.}\ }\href {https://doi.org/10.1016/j.apradiso.2022.110905}
  {10.1016/j.apradiso.2022.110905} (\bibinfo {year} {2023})\BibitemShut
  {NoStop}%
\bibitem [{\citenamefont {Hindi{\'e}}\ and\ \citenamefont
  {et~al.}(2024)}]{Hindie2024_Pd103_Dosimetry}%
  \BibitemOpen
  \bibfield  {author} {\bibinfo {author} {\bibfnamefont {E.}~\bibnamefont
  {Hindi{\'e}}}\ and\ \bibinfo {author} {\bibnamefont {et~al.}},\ }\bibfield
  {title} {\bibinfo {title} {Absorbed doses in single cells for
  $^{103}$pd(/$^{103m}$rh)},\ }\href
  {https://pmc.ncbi.nlm.nih.gov/articles/PMC11303077/} {\bibfield  {journal}
  {\bibinfo  {journal} {EJNMMI Phys.}\ } (\bibinfo {year} {2024})}\BibitemShut
  {NoStop}%
\bibitem [{\citenamefont {Harper}\ \emph {et~al.}(1965)\citenamefont {Harper},
  \citenamefont {Lathrop}, \citenamefont {Jiminez}, \citenamefont {Fink},\ and\
  \citenamefont {Gottschalk}}]{harper1965technetium}%
  \BibitemOpen
  \bibfield  {author} {\bibinfo {author} {\bibfnamefont {P.}~\bibnamefont
  {Harper}}, \bibinfo {author} {\bibfnamefont {K.}~\bibnamefont {Lathrop}},
  \bibinfo {author} {\bibfnamefont {F.}~\bibnamefont {Jiminez}}, \bibinfo
  {author} {\bibfnamefont {R.}~\bibnamefont {Fink}},\ and\ \bibinfo {author}
  {\bibfnamefont {A.}~\bibnamefont {Gottschalk}},\ }\bibfield  {title}
  {\bibinfo {title} {Technetium 99m as a scanning agent},\ }\href@noop {}
  {\bibfield  {journal} {\bibinfo  {journal} {Radiology}\ }\textbf {\bibinfo
  {volume} {85}},\ \bibinfo {pages} {101} (\bibinfo {year} {1965})}\BibitemShut
  {NoStop}%
\bibitem [{\citenamefont {Eckelman}(2009)}]{eckelman2009unparalleled}%
  \BibitemOpen
  \bibfield  {author} {\bibinfo {author} {\bibfnamefont {W.~C.}\ \bibnamefont
  {Eckelman}},\ }\bibfield  {title} {\bibinfo {title} {Unparalleled
  contribution of technetium-99m to medicine over 5 decades},\ }\href@noop {}
  {\bibfield  {journal} {\bibinfo  {journal} {JACC: Cardiovascular Imaging}\
  }\textbf {\bibinfo {volume} {2}},\ \bibinfo {pages} {364} (\bibinfo {year}
  {2009})}\BibitemShut {NoStop}%
\bibitem [{\citenamefont
  {Papagiannopoulou}(2017)}]{papagiannopoulou2017technetium}%
  \BibitemOpen
  \bibfield  {author} {\bibinfo {author} {\bibfnamefont {D.}~\bibnamefont
  {Papagiannopoulou}},\ }\bibfield  {title} {\bibinfo {title} {Technetium-99m
  radiochemistry for pharmaceutical applications},\ }\href@noop {} {\bibfield
  {journal} {\bibinfo  {journal} {Journal of Labelled Compounds and
  Radiopharmaceuticals}\ }\textbf {\bibinfo {volume} {60}},\ \bibinfo {pages}
  {502} (\bibinfo {year} {2017})}\BibitemShut {NoStop}%
\bibitem [{\citenamefont {Gascoine}(2021)}]{gascoine2021towards}%
  \BibitemOpen
  \bibfield  {author} {\bibinfo {author} {\bibfnamefont {M.}~\bibnamefont
  {Gascoine}},\ }\bibfield  {title} {\bibinfo {title} {Towards the fast
  neutron-induced isotope production of 99mtc via the 102ru (n, $\alpha$) 99mo
  reaction.},\ }\href@noop {} {\  (\bibinfo {year} {2021})}\BibitemShut
  {NoStop}%
\bibitem [{\citenamefont {Poty}\ \emph
  {et~al.}(2018{\natexlab{a}})\citenamefont {Poty}, \citenamefont
  {Francesconi}, \citenamefont {McDevitt}, \citenamefont {Morris},\ and\
  \citenamefont {Lewis}}]{poty2018alpha}%
  \BibitemOpen
  \bibfield  {author} {\bibinfo {author} {\bibfnamefont {S.}~\bibnamefont
  {Poty}}, \bibinfo {author} {\bibfnamefont {L.~C.}\ \bibnamefont
  {Francesconi}}, \bibinfo {author} {\bibfnamefont {M.~R.}\ \bibnamefont
  {McDevitt}}, \bibinfo {author} {\bibfnamefont {M.~J.}\ \bibnamefont
  {Morris}},\ and\ \bibinfo {author} {\bibfnamefont {J.~S.}\ \bibnamefont
  {Lewis}},\ }\bibfield  {title} {\bibinfo {title} {$\alpha$-emitters for
  radiotherapy: from basic radiochemistry to clinical studies---part 1},\
  }\href@noop {} {\bibfield  {journal} {\bibinfo  {journal} {Journal of Nuclear
  Medicine}\ }\textbf {\bibinfo {volume} {59}},\ \bibinfo {pages} {878}
  (\bibinfo {year} {2018}{\natexlab{a}})}\BibitemShut {NoStop}%
\bibitem [{\citenamefont {Poty}\ \emph
  {et~al.}(2018{\natexlab{b}})\citenamefont {Poty}, \citenamefont
  {Francesconi}, \citenamefont {McDevitt}, \citenamefont {Morris},\ and\
  \citenamefont {Lewis}}]{poty2018alphapt2}%
  \BibitemOpen
  \bibfield  {author} {\bibinfo {author} {\bibfnamefont {S.}~\bibnamefont
  {Poty}}, \bibinfo {author} {\bibfnamefont {L.~C.}\ \bibnamefont
  {Francesconi}}, \bibinfo {author} {\bibfnamefont {M.~R.}\ \bibnamefont
  {McDevitt}}, \bibinfo {author} {\bibfnamefont {M.~J.}\ \bibnamefont
  {Morris}},\ and\ \bibinfo {author} {\bibfnamefont {J.~S.}\ \bibnamefont
  {Lewis}},\ }\bibfield  {title} {\bibinfo {title} {$\alpha$-emitters for
  radiotherapy: from basic radiochemistry to clinical studies---part 2},\
  }\href@noop {} {\bibfield  {journal} {\bibinfo  {journal} {Journal of Nuclear
  Medicine}\ }\textbf {\bibinfo {volume} {59}},\ \bibinfo {pages} {1020}
  (\bibinfo {year} {2018}{\natexlab{b}})}\BibitemShut {NoStop}%
\bibitem [{\citenamefont {Iyengar}(1990)}]{iyengar1990natural}%
  \BibitemOpen
  \bibfield  {author} {\bibinfo {author} {\bibfnamefont {M.}~\bibnamefont
  {Iyengar}},\ }\bibfield  {title} {\bibinfo {title} {The natural distribution
  of radium},\ }\href@noop {} {\bibfield  {journal} {\bibinfo  {journal} {The
  environmental behaviour of radium}\ }\textbf {\bibinfo {volume} {1}},\
  \bibinfo {pages} {59} (\bibinfo {year} {1990})}\BibitemShut {NoStop}%
\bibitem [{\citenamefont {Bateman}(1910)}]{bateman1910solution}%
  \BibitemOpen
  \bibfield  {author} {\bibinfo {author} {\bibfnamefont {H.}~\bibnamefont
  {Bateman}},\ }\bibfield  {title} {\bibinfo {title} {The solution of a system
  of differential equations occurring in the theory of radioactive
  transformations},\ }in\ \href@noop {} {\emph {\bibinfo {booktitle} {Proc.
  Cambridge Philos. Soc.}}},\ Vol.~\bibinfo {volume} {15}\ (\bibinfo {year}
  {1910})\ pp.\ \bibinfo {pages} {423--427}\BibitemShut {NoStop}%
\bibitem [{\citenamefont {Cetnar}(2006)}]{cetnar2006general}%
  \BibitemOpen
  \bibfield  {author} {\bibinfo {author} {\bibfnamefont {J.}~\bibnamefont
  {Cetnar}},\ }\bibfield  {title} {\bibinfo {title} {General solution of
  bateman equations for nuclear transmutations},\ }\href@noop {} {\bibfield
  {journal} {\bibinfo  {journal} {Annals of Nuclear Energy}\ }\textbf {\bibinfo
  {volume} {33}},\ \bibinfo {pages} {640} (\bibinfo {year} {2006})}\BibitemShut
  {NoStop}%
\bibitem [{\citenamefont {Kovari}\ \emph {et~al.}(2017)\citenamefont {Kovari},
  \citenamefont {Coleman}, \citenamefont {Cristescu},\ and\ \citenamefont
  {Smith}}]{kovari2017tritium}%
  \BibitemOpen
  \bibfield  {author} {\bibinfo {author} {\bibfnamefont {M.}~\bibnamefont
  {Kovari}}, \bibinfo {author} {\bibfnamefont {M.}~\bibnamefont {Coleman}},
  \bibinfo {author} {\bibfnamefont {I.}~\bibnamefont {Cristescu}},\ and\
  \bibinfo {author} {\bibfnamefont {R.}~\bibnamefont {Smith}},\ }\bibfield
  {title} {\bibinfo {title} {Tritium resources available for fusion reactors},\
  }\href@noop {} {\bibfield  {journal} {\bibinfo  {journal} {Nuclear Fusion}\
  }\textbf {\bibinfo {volume} {58}},\ \bibinfo {pages} {026010} (\bibinfo
  {year} {2017})}\BibitemShut {NoStop}%
\bibitem [{\citenamefont {Abdou}(1995)}]{abdou1995volumetric}%
  \BibitemOpen
  \bibfield  {author} {\bibinfo {author} {\bibfnamefont {M.~A.}\ \bibnamefont
  {Abdou}},\ }\bibfield  {title} {\bibinfo {title} {A volumetric neutron source
  for fusion nuclear technology testing and development},\ }\href@noop {}
  {\bibfield  {journal} {\bibinfo  {journal} {Fusion engineering and design}\
  }\textbf {\bibinfo {volume} {27}},\ \bibinfo {pages} {111} (\bibinfo {year}
  {1995})}\BibitemShut {NoStop}%
\bibitem [{\citenamefont {Kuteev}\ \emph {et~al.}(2010)\citenamefont {Kuteev},
  \citenamefont {Goncharov}, \citenamefont {Sergeev},\ and\ \citenamefont
  {Khripunov}}]{kuteev2010intense}%
  \BibitemOpen
  \bibfield  {author} {\bibinfo {author} {\bibfnamefont {B.}~\bibnamefont
  {Kuteev}}, \bibinfo {author} {\bibfnamefont {P.}~\bibnamefont {Goncharov}},
  \bibinfo {author} {\bibfnamefont {V.~Y.}\ \bibnamefont {Sergeev}},\ and\
  \bibinfo {author} {\bibfnamefont {V.}~\bibnamefont {Khripunov}},\ }\bibfield
  {title} {\bibinfo {title} {Intense fusion neutron sources},\ }\href@noop {}
  {\bibfield  {journal} {\bibinfo  {journal} {Plasma Physics Reports}\ }\textbf
  {\bibinfo {volume} {36}},\ \bibinfo {pages} {281} (\bibinfo {year}
  {2010})}\BibitemShut {NoStop}%
\bibitem [{\citenamefont {Kr{\'o}las}\ \emph {et~al.}(2021)\citenamefont
  {Kr{\'o}las}, \citenamefont {Ibarra}, \citenamefont {Arbeiter}, \citenamefont
  {Arranz}, \citenamefont {Bernardi}, \citenamefont {Cappelli}, \citenamefont
  {Castellanos}, \citenamefont {D{\'e}zsi}, \citenamefont {Dzitko},
  \citenamefont {Favuzza} \emph {et~al.}}]{krolas2021ifmif}%
  \BibitemOpen
  \bibfield  {author} {\bibinfo {author} {\bibfnamefont {W.}~\bibnamefont
  {Kr{\'o}las}}, \bibinfo {author} {\bibfnamefont {A.}~\bibnamefont {Ibarra}},
  \bibinfo {author} {\bibfnamefont {F.}~\bibnamefont {Arbeiter}}, \bibinfo
  {author} {\bibfnamefont {F.}~\bibnamefont {Arranz}}, \bibinfo {author}
  {\bibfnamefont {D.}~\bibnamefont {Bernardi}}, \bibinfo {author}
  {\bibfnamefont {M.}~\bibnamefont {Cappelli}}, \bibinfo {author}
  {\bibfnamefont {J.}~\bibnamefont {Castellanos}}, \bibinfo {author}
  {\bibfnamefont {T.}~\bibnamefont {D{\'e}zsi}}, \bibinfo {author}
  {\bibfnamefont {H.}~\bibnamefont {Dzitko}}, \bibinfo {author} {\bibfnamefont
  {P.}~\bibnamefont {Favuzza}}, \emph {et~al.},\ }\bibfield  {title} {\bibinfo
  {title} {The ifmif-dones fusion oriented neutron source: evolution of the
  design},\ }\href@noop {} {\bibfield  {journal} {\bibinfo  {journal} {Nuclear
  Fusion}\ }\textbf {\bibinfo {volume} {61}},\ \bibinfo {pages} {125002}
  (\bibinfo {year} {2021})}\BibitemShut {NoStop}%
\bibitem [{\citenamefont {Giannini}\ \emph {et~al.}(2024)\citenamefont
  {Giannini}, \citenamefont {Luongo}, \citenamefont {Federici}, \citenamefont
  {Bachmann}, \citenamefont {Siccinio},\ and\ \citenamefont
  {Leichtle}}]{giannini2024conceptual}%
  \BibitemOpen
  \bibfield  {author} {\bibinfo {author} {\bibfnamefont {L.}~\bibnamefont
  {Giannini}}, \bibinfo {author} {\bibfnamefont {C.}~\bibnamefont {Luongo}},
  \bibinfo {author} {\bibfnamefont {G.}~\bibnamefont {Federici}}, \bibinfo
  {author} {\bibfnamefont {C.}~\bibnamefont {Bachmann}}, \bibinfo {author}
  {\bibfnamefont {M.}~\bibnamefont {Siccinio}},\ and\ \bibinfo {author}
  {\bibfnamefont {D.}~\bibnamefont {Leichtle}},\ }\bibfield  {title} {\bibinfo
  {title} {Conceptual design studies on the magnet system for the volumetric
  neutron source},\ }\href@noop {} {\bibfield  {journal} {\bibinfo  {journal}
  {IEEE Transactions on Applied Superconductivity}\ }\textbf {\bibinfo {volume}
  {34}},\ \bibinfo {pages} {1} (\bibinfo {year} {2024})}\BibitemShut {NoStop}%
\bibitem [{\citenamefont {Vigan{\`o}}\ \emph {et~al.}(2025)\citenamefont
  {Vigan{\`o}}, \citenamefont {Pagani}, \citenamefont {Talloni}, \citenamefont
  {Haghdoust}, \citenamefont {Falcitelli}, \citenamefont {Maione},
  \citenamefont {Giannini}, \citenamefont {Luongo},\ and\ \citenamefont
  {Lucca}}]{vigano2025multidisciplinary}%
  \BibitemOpen
  \bibfield  {author} {\bibinfo {author} {\bibfnamefont {F.}~\bibnamefont
  {Vigan{\`o}}}, \bibinfo {author} {\bibfnamefont {I.}~\bibnamefont {Pagani}},
  \bibinfo {author} {\bibfnamefont {S.}~\bibnamefont {Talloni}}, \bibinfo
  {author} {\bibfnamefont {P.}~\bibnamefont {Haghdoust}}, \bibinfo {author}
  {\bibfnamefont {G.}~\bibnamefont {Falcitelli}}, \bibinfo {author}
  {\bibfnamefont {I.}~\bibnamefont {Maione}}, \bibinfo {author} {\bibfnamefont
  {L.}~\bibnamefont {Giannini}}, \bibinfo {author} {\bibfnamefont
  {C.}~\bibnamefont {Luongo}},\ and\ \bibinfo {author} {\bibfnamefont
  {F.}~\bibnamefont {Lucca}},\ }\bibfield  {title} {\bibinfo {title} {A
  multidisciplinary approach to volumetric neutron source (vns) thermal shield
  design: Analysis and optimisation of electromagnetic, thermal, and structural
  behaviours},\ }\href@noop {} {\bibfield  {journal} {\bibinfo  {journal}
  {Energies}\ }\textbf {\bibinfo {volume} {18}},\ \bibinfo {pages} {3305}
  (\bibinfo {year} {2025})}\BibitemShut {NoStop}%
\bibitem [{\citenamefont {Hendel}\ and\ \citenamefont
  {Jassby}(1990)}]{Hendel01101990}%
  \BibitemOpen
  \bibfield  {author} {\bibinfo {author} {\bibfnamefont {H.~W.}\ \bibnamefont
  {Hendel}}\ and\ \bibinfo {author} {\bibfnamefont {D.~L.}\ \bibnamefont
  {Jassby}},\ }\bibfield  {title} {\bibinfo {title} {The tokamak as a neutron
  source},\ }\href {https://doi.org/10.13182/NSE90-A27465} {\bibfield
  {journal} {\bibinfo  {journal} {Nuclear Science and Engineering}\ }\textbf
  {\bibinfo {volume} {106}},\ \bibinfo {pages} {114} (\bibinfo {year}
  {1990})},\ \Eprint
  {https://arxiv.org/abs/https://doi.org/10.13182/NSE90-A27465}
  {https://doi.org/10.13182/NSE90-A27465} \BibitemShut {NoStop}%
\bibitem [{\citenamefont {Swanson}\ \emph {et~al.}(2025)\citenamefont
  {Swanson}, \citenamefont {Gates}, \citenamefont {Kumar}, \citenamefont
  {Martin}, \citenamefont {Kruger}, \citenamefont {Dudt}, \citenamefont
  {Bonofiglo} \emph {et~al.}}]{swanson2025scoping}%
  \BibitemOpen
  \bibfield  {author} {\bibinfo {author} {\bibfnamefont {C.}~\bibnamefont
  {Swanson}}, \bibinfo {author} {\bibfnamefont {D.}~\bibnamefont {Gates}},
  \bibinfo {author} {\bibfnamefont {S.}~\bibnamefont {Kumar}}, \bibinfo
  {author} {\bibfnamefont {M.}~\bibnamefont {Martin}}, \bibinfo {author}
  {\bibfnamefont {T.}~\bibnamefont {Kruger}}, \bibinfo {author} {\bibfnamefont
  {D.}~\bibnamefont {Dudt}}, \bibinfo {author} {\bibfnamefont {P.}~\bibnamefont
  {Bonofiglo}}, \emph {et~al.},\ }\bibfield  {title} {\bibinfo {title} {The
  scoping, design, and plasma physics optimization of the eos neutron source
  stellarator},\ }\href@noop {} {\bibfield  {journal} {\bibinfo  {journal}
  {Nuclear Fusion}\ }\textbf {\bibinfo {volume} {65}},\ \bibinfo {pages}
  {026053} (\bibinfo {year} {2025})}\BibitemShut {NoStop}%
\end{thebibliography}%

\end{document}